\documentclass{aa}
\usepackage{natbib}
\bibpunct{(}{)}{;}{a}{}{,}
\usepackage{graphicx}
\usepackage{txfonts}
\begin{document}
   \title{Eclipsed X-ray flares in binary stars: geometrical
     constraints on the flare's location and size
     \thanks{Figs. \ref{vwcepres}--\ref{simulations} and
       \ref{fig:examplearcresults}--\ref{fig:test3} are only available
       in electronic form via http://www.edpsciences.org}
}

   \author{J. Sanz-Forcada\inst{1}
     \thanks{\emph{Present address:}
     Laboratorio de Astrof\'{i}sica Espacial y F\'{i}sica Fundamental, INTA, P.O. Box 50727, E-28080 Madrid (Spain)}
          \and
          F. Favata\inst{1}
          \and
          G. Micela\inst{2}
          }

   \offprints{J. Sanz-Forcada, \email{jsanz@laeff.inta.es}}

   \institute{Astrophysics Division -- Research and Science Support Department
  	of ESA, ESTEC, Postbus 299, NL-2200 AG Noordwijk, The Netherlands\\
              \email{jsanz@laeff.inta.es, ffavata@rssd.esa.int}
         \and
             INAF - Osservatorio Astronomico di Palermo,
             Piazza del Parlamento 1, I-90134 Palermo, Italy
             \email{giusi@astropa.unipa.it}
             }

   \date{Received / Accepted 9 January 2007}

   \abstract{}
   {The observation of eclipses during X-rays flares taking place in
     active cool stars binaries  allows us to
     calculate the position and size of the flares. 
     This information cannot be
     derived by analyzing the decay of the flares, a frequently used
     approach in the 
     literature that requires the assumption of a physical model. We
     make use of the  
     eclipsing light curve to constrain the set of possible solutions,
     from the geometrical point of view, in two flares of Algol, and
     one flare in VW Cep.}
   {We make use of a technique developed with the system SV~Cam
     ($i\sim 90\degr$) and generalize it to binary systems with arbitrary
     inclination. The method simulates all possible geometrical
     situations that can produce the times of the four contacts of
     the eclipse. As an approximation we assume that the emitting region
     has a spherical shape that remains unchanged during the eclipse.
     We however show that this is a good approximation for the problem.}
   {The solutions observed indicate that in two of the three cases the
     flare cannot be polar ($|\theta| < 55\degr$) and in a third one
     the flare can be placed either near the pole or at other
     latitudes. The emitting regions must have a small size ($0.002 -
     0.5 R_*$), but if interpreted as the apex of
     coronal loops, their length could actually be up to 3.1~$R_*$
     for one of the Algol flares. These measurements imply a  
   lower limit to the electron density in the emitting region between $\log
   n_e$\,(cm$^{-3}$) 10.4 and 14.0, and a magnetic field between 70
   and 3500~G. Similar results are found if the emitting region
     is assumed to be loop-shaped.
}
   {}

   \maketitle
%

\section{Introduction}
The determination of the dimension of the X-ray emitting region in stellar flares has in most cases been based on assumptions on the
physics of the flaring emission, in particular the rise and decay
times. 
Although it is not possible to directly observe spatially resolved flares
in stars (except for the Sun), it is possible to constrain their
geometry if the flare is eclipsed by one of the stars in the system. 

There are four eclipsed X-ray flares reported in the
literature: one in VW~Cep \citep{cho98},
two in Algol \citep{schm99,sch03} and very recently one in SV~Cam
\citep[][hereafter Paper I]{sanz06}. In Paper I we calculated all the
possible solutions for the eclipse of SV~Cam, a system with $i\sim 90\degr$,
which makes the geometrical problem simpler. We extend in this work
the methodology applied in Paper I to the other three eclipsed
flares, which implies the use of the
equations for the general case $i\neq 90\degr$. 

\begin{figure}
   \centering
   \includegraphics[angle=90,width=0.45\textwidth]{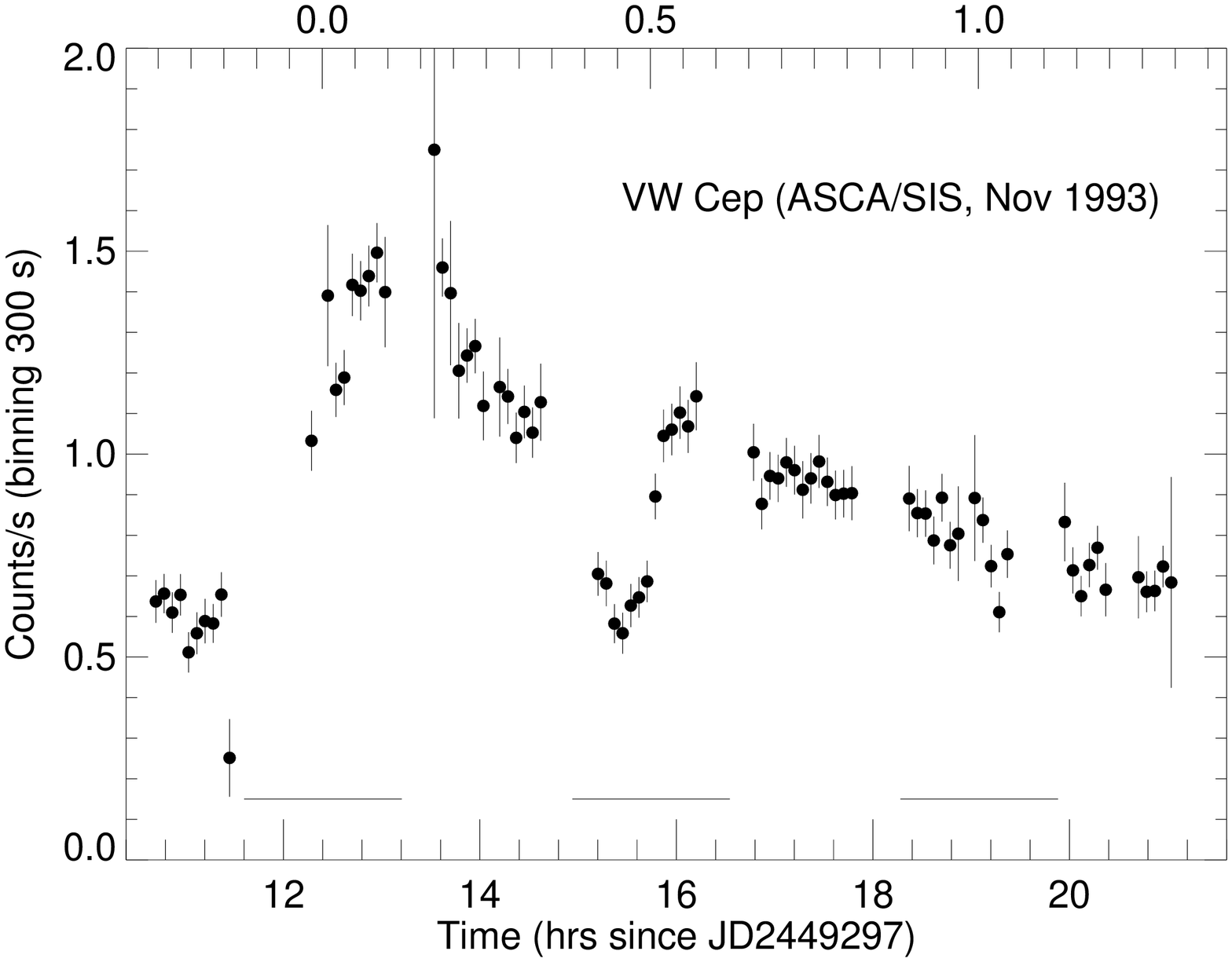}
   \includegraphics[angle=90,width=0.45\textwidth]{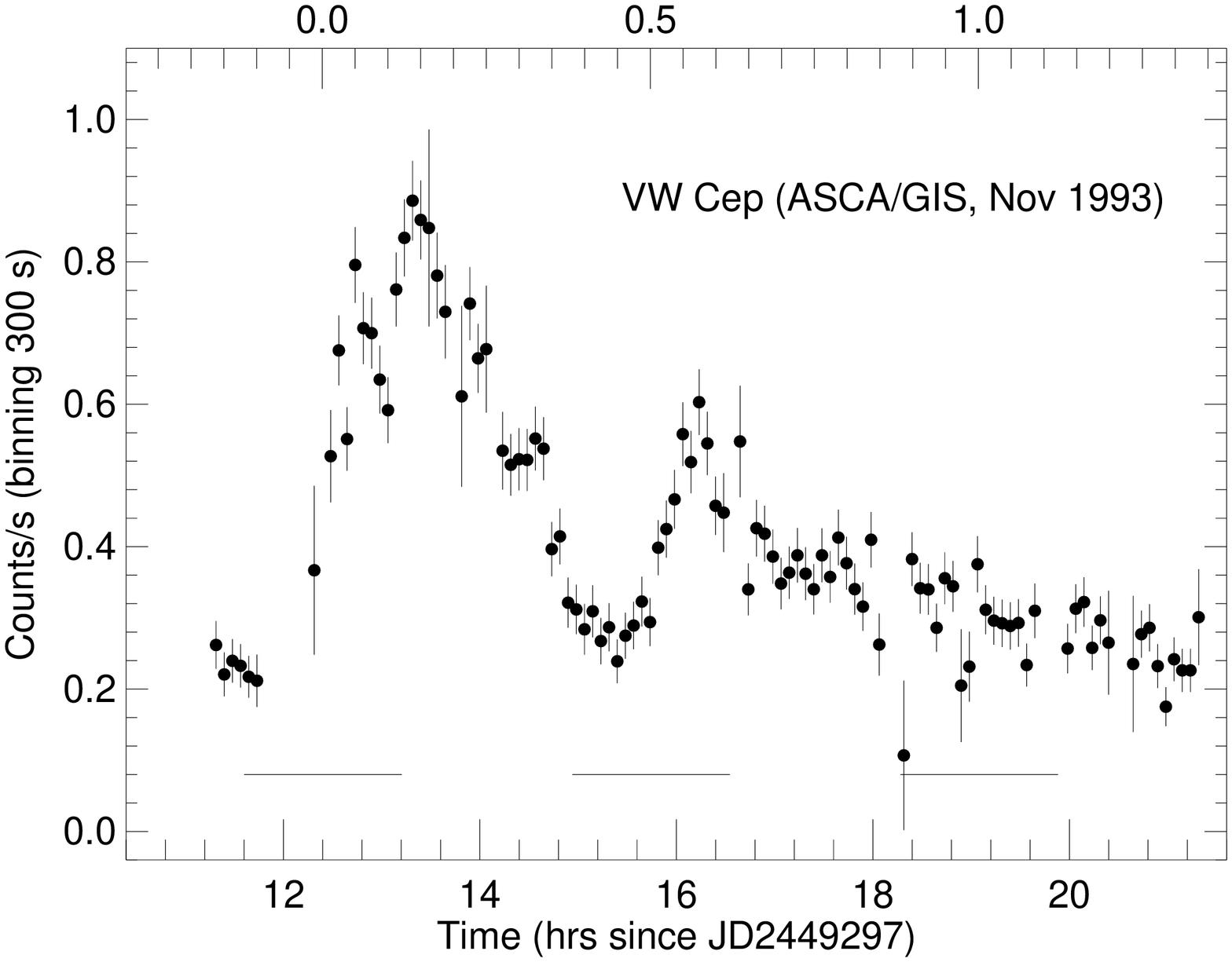}
   \includegraphics[angle=90,width=0.45\textwidth]{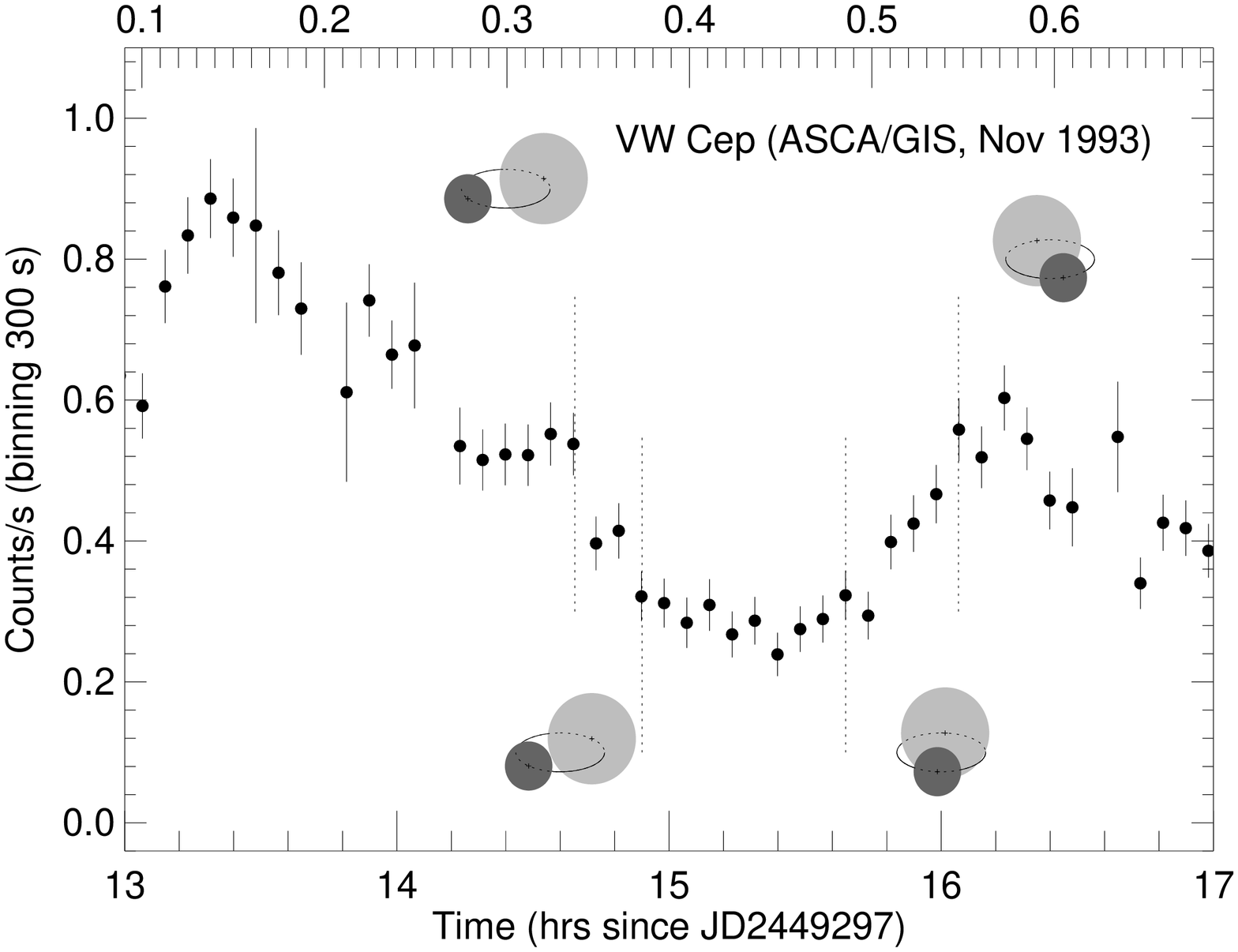}
   \caption{VW Cep X-ray light curve (ASCA/SIS and ASCA/GIS) in
     Feb. 2002. The solid lines mark the times of the optical eclipses. An
     eclipse in the flare, around $\phi \sim 0.45$, is displayed in
     the lower panel, where dotted lines indicate the phases of the 
     four contacts
     used in the analysis, with a representation of the relative
     postions of the stars next to them.}
   \label{vwceplc}
\end{figure}

The first case reported in the literature of an eclipse of a flare in
X-rays was an event in VW Cep (G5V/K0V), 
a binary system of the W UMa-type. These are eclipsing binaries that 
consist in F-K main sequence stars with short
orbital periods. They are, in general, strong X-ray sources. 
The observed X-ray emission
is attributed to magnetic activity caused by a dynamo. 
VW~Cep is an unusual system, since the primary star (the
deepest eclipse, at orbital phase $\phi$=0, occurs when this star is
occulted, with $i=65~\degr$) is the hotter (G5V) but the less massive
star ($M=0.25\,M_\odot$,  
$R=0.50\,R_\odot$), while the secondary (K0V, $M=0.90\,M_\odot$,
$R=0.93\,R_\odot$) is more massive \citep[see][and references
therein]{hil89,cho98}. 
\citet{cho98} analyzed the {\em ASCA} light curve of VW~Cep proposing
that the observed dip during the decay of a flare, at $\phi \sim
0.42$ (Fig. \ref{vwceplc}), was produced by the eclipse of this
flare. From investigation of the light curve they also conclude that
the flare took place in the 
polar region of the most massive star (the K0V).

Algol (B8V/K2III) is an interesting target to search for eclipses
($i=82.5~\degr$) in
X-rays, since the hot component of the system can be safely assumed to
be dark in X-rays as compared to the K2III secondary, which is expected to host
coronal features. Two eclipsed flares have been reported in
Algol. \citet{schm99} observed, in 1997, an eclipse with {\em Beppo-SAX} during
the decay of a large flare. They estimated that the flare reached a
maximum of 0.6\,$R_*$ above the stellar surface, in
a polar region in order to produce the observed eclipse centered around $\phi \simeq 0.50$, and a density of
at least $\sim 10^{11}$\,cm$^{-3}$ was necessary.
\citet{sch03} reported another eclipse during a flare in Algol,
observed with {\em XMM-Newton} in 2001, and they used a reconstruction method
to conclude that the flare had a height of $\sim 0.1R_*$
(0.3\,$R_\odot$) over the surface ($R_{1,2}=3.09/3.29R_\odot$), placed near
the limb, with a density above $\sim 10^{11}$\,cm$^{-3}$. 

\begin{figure}
   \centering
   \includegraphics[angle=90,width=0.45\textwidth]{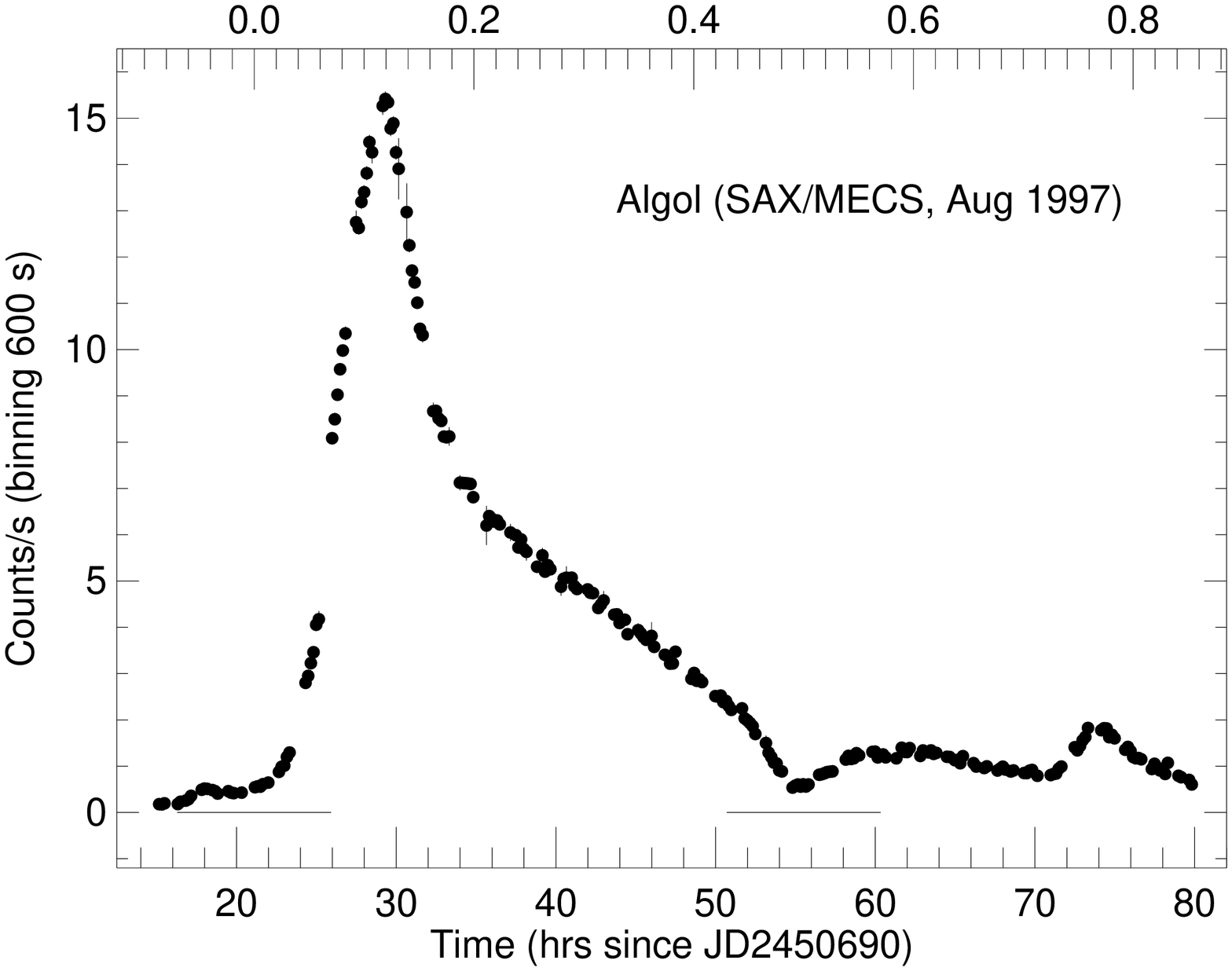}
   \includegraphics[angle=90,width=0.45\textwidth]{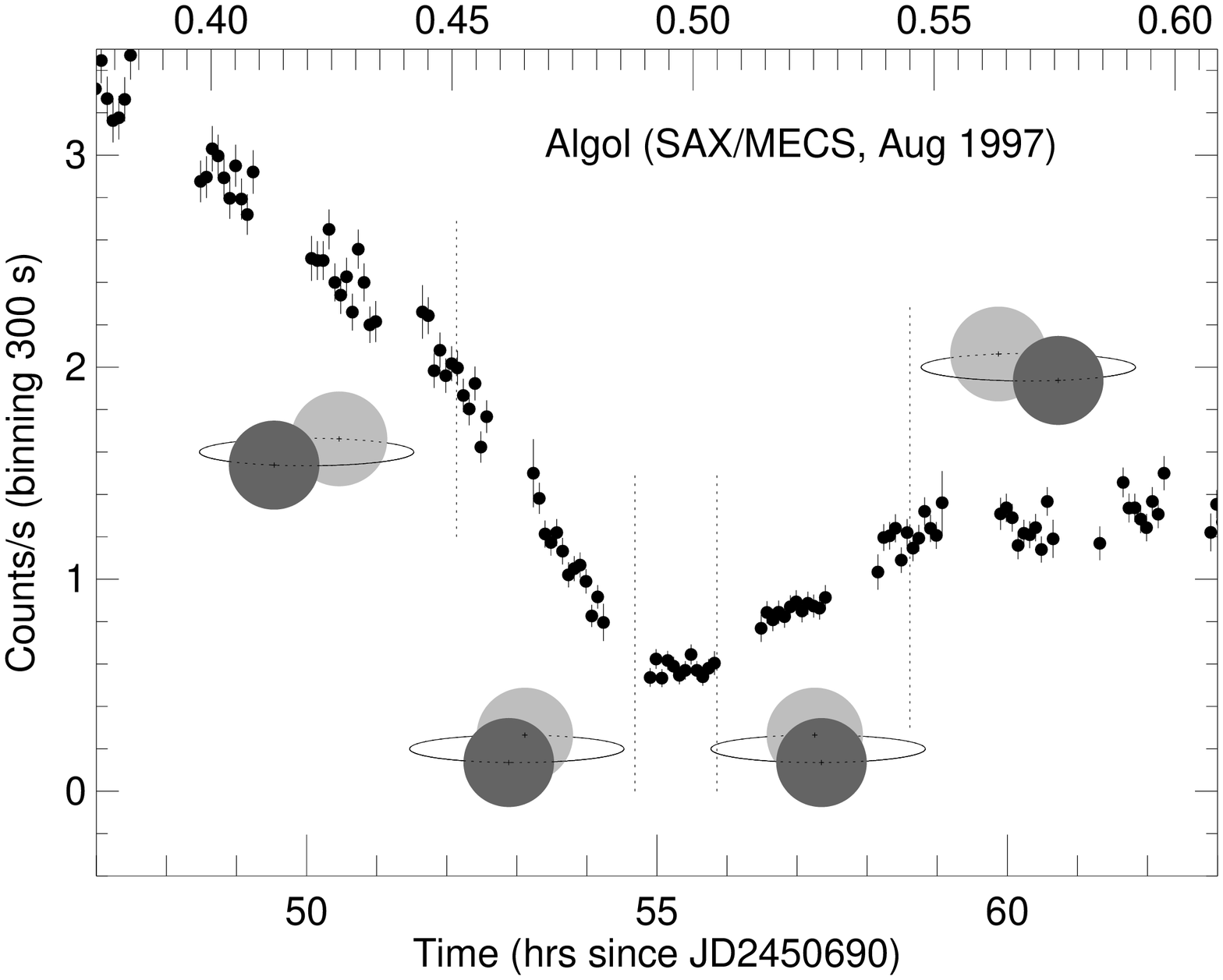}
   \caption{Algol light curve (BeppoSAX/MECS) in Aug 1997. An eclipse
   in the flare, around $\phi \sim 0.5$, is enlarged in the lower panel.}
   \label{algol1lc}
\end{figure}

In this work we apply the same technique already used with SV~Cam
(Paper I) to the three cases mentioned above. In the case of VW~Cep
and the XMM observation of Algol we reanalyzed the data to calculate
the phases of the four contacts. The paper is structured as follows:
in Sect.~2 we describe the technical details of the observations;
Sect.~3 develops the geometrical formulae used in this technique and
the results found for each case. The results and their implications in
the coronal context are discussed in Sect.~4, and the conclusions are
given in Sect.~5.

\section{Observations}
VW Cep (HD 197433) was observed with ASCA on 5 Nov. 1993 23:37 UT
\citep{cho98}. The ASCA 
satellite has four telescopes equipped with two Solid-State Imaging
Spectrometers (SIS0, SIS1, range 0.4--10~keV), and two Gas Imaging
Spectrometers (GIS2, GIS3, range 0.8--10~keV). We have reanalyzed the
light curves by selecting a circular region around VW~Cep, and
subtracting the background level. The light curve displayed in
Fig.~\ref{vwceplc} has been obtained by correcting for small time
shifts between the two GIS datasets and then summing the signal. Times
corresponding to the four contacts were determined in the GIS summed
light curve. 

\begin{figure}
   \centering
   \includegraphics[angle=90,width=0.45\textwidth]{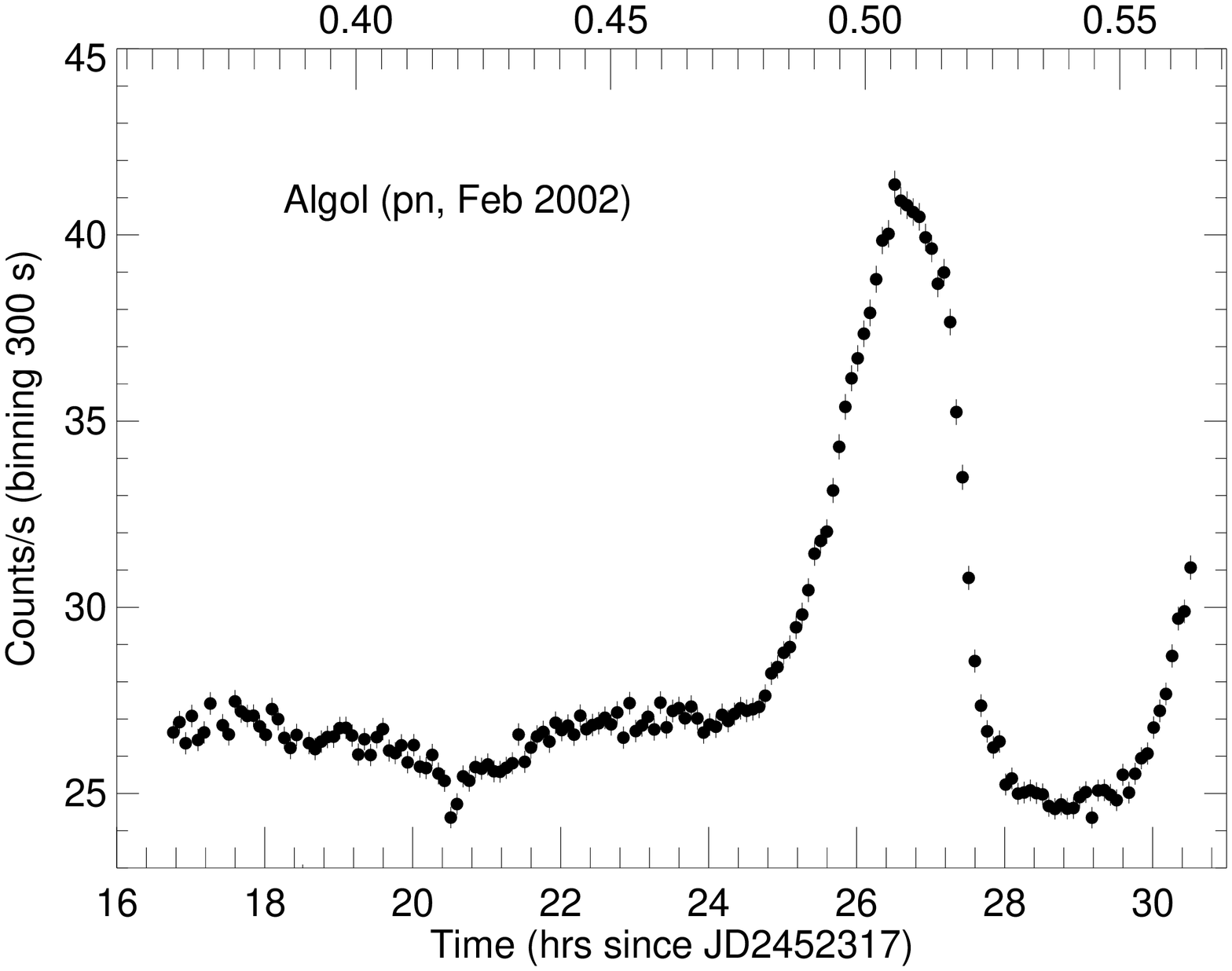}
   \includegraphics[angle=90,width=0.45\textwidth]{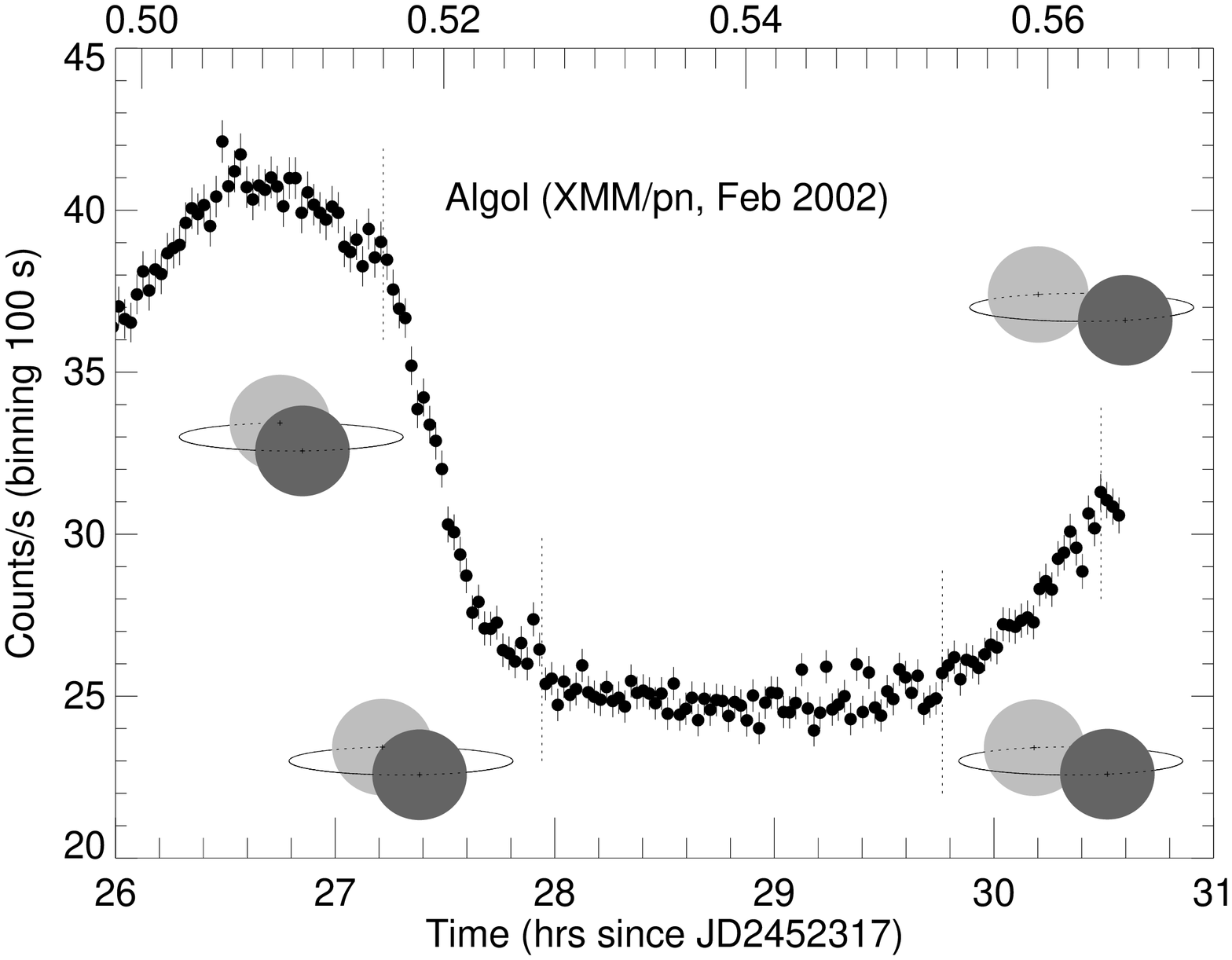}
   \caption{Algol light curve (EPIC/PN) in Feb 2002. The optical
  eclipse takes place between $\phi\sim 0.43$ and $\phi\sim 0.57$.
  The eclipse in the flare, around $\phi \sim 0.54$, is displayed
  in the lower panel.}
   \label{algol2lc}
\end{figure}

Two observations of Algol ($\beta$~Per, HD 19356, HR 936) are analyzed
in this work. Algol was observed on 30 Aug. 1997 3:04 UT with BeppoSAX
\citep{schm99}.  We use in this case the phases of the four contacts
as provided by the authors. The corresponding light curve
(Fig.~\ref{algol1lc}), was
constructed summing the data from all the three  Medium Energy
Concentrator Spectrometer (MECS) detectors (range 0.1-10~keV).

{\em XMM-Newton} observed Algol on 12 Feb. 2002 4:42 UT
\citep{sch03}. The light curve used for the analysis is based on
EPIC-PN (range 0.15-15~keV) data that we reanalyzed using the
standard SAS (Science Analysis Software) version 6.1. We determined
the four contacts of the eclipse as displayed in Fig.~\ref{algol2lc}.

\section{Results}

\begin{figure}
   \centering
   \includegraphics[width=0.50\textwidth]{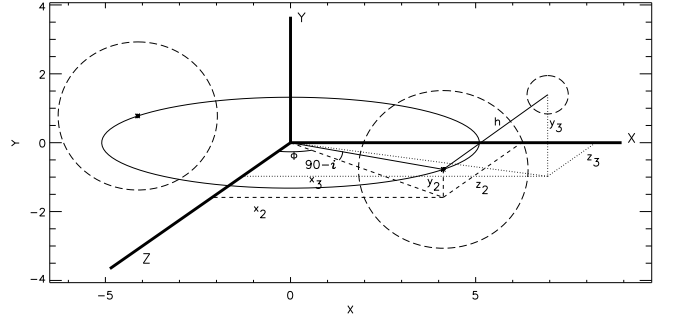}
   \caption{Schematic view of the geometry of the stars and the
     flaring region (dashed) at phase $\phi \sim 0.15$.}
   \label{fig:eclipse}
\end{figure}

In Paper I we made an analysis of an eclipse of a flare in
SV~Cam. In that case the high inclination ($i \sim 90\degr$) of the
system simplifies the equations. Here we calculate new equations for the general case $i \neq 90\degr$. 
We assume that the emitting region has a spherical shape, and is defined by the following quantities: latitude ($\theta$), longitude 
($\lambda$, with origin 0 the meridian of each star that
is in front of the observer at $\phi$=0), height over the center 
of the star ($h$) and the radius of the spherical emitting region ($R_3$).
We have made a grid of values for the 4 variables, and we then calculated
the times when the four contacts of the eclipse take place for each set of
values. We considered as valid results those that agree within 1 time
bin (300~s for VW~Cep and the Algol XMM observation, 600~s for
Algol with BeppoSAX) with the measured times.
We describe here the equations corresponding to a system with a
primary star that hosts a flare which is eclipsed by the secondary.
For the simple case (system with $i=90\degr$):

{\setlength\arraycolsep{1pt}
\begin{eqnarray*}
x'_1 & = & - a \sin(\phi), \quad y'_1=0, \quad
z'_1=- a \cos{\phi} \\
x'_2 & = & -x'_1, \quad y'_2=0, \quad z'_2=-z'_1 \\
x'_3 & = & x'_1+h \cos(\theta) \sin(\phi+\lambda), \quad 
y'_3 = y'_1+h \sin(\theta), \quad  \\
z'_3 & = & z'_1+h \cos(\theta) \cos(\phi+\lambda)
\end{eqnarray*}}
where $x'$, $y'$ correspond to the coordinates in the 
plane in front of the observer
(Fig.~\ref{fig:eclipse}), $z'$ is the coordinate of the axis
perpendicular to this plane 
(positive towards the observer), and $a$ is the value of the semi-axis
of the orbit. $x'_3$, $y'_3$, $z'_3$ correspond to the 
position of the emitting region of the flare, and the subindices 1 and
2 correspond to the positions of the centers of the primary and secondary stars
respectively. 

Now, for the general case in which $i\neq 90\degr $ we need to apply a
rotation. In the new coordinates system, $x, y, z$ can be calculated
from the former coordinates: 

\begin{displaymath}
x=x', \quad y=\sin i \, y' - \cos i \, z', \quad z=\cos i \, y' + \sin
i \, z'
\end{displaymath}
and therefore,
{\setlength\arraycolsep{1pt}
\begin{eqnarray*}
x_1 & = & -a \sin \phi, \quad y_1=a \cos \phi \cos i, \quad
z_1=-a \cos{\phi} \sin i  \\
x_2 & = & a \sin \phi, \quad y_2=-a \cos \phi \cos i, \quad 
z_2=a \cos{\phi} \sin i \\
x_3 & = & x_1 + h \cos \theta \sin(\phi+\lambda), \quad \\
y_3 & = & y_1 + h [\sin \theta \sin i - \cos \theta \cos(\phi+\lambda)
\cos i ] \\
z_3 & = & z_1 + h [\sin \theta \cos i + \cos \theta \cos(\phi+\lambda)
\sin i ]
\end{eqnarray*}}

\begin{table*}
\caption{VW Cep (G5V/K0V, $R_{1,2}$=0.50/0.93~R$_\odot$) range of possible
solutions of the eclipse of a flare in
1993.}\label{tab:vwcep} 
\begin{center}
\begin{footnotesize}
 \begin{tabular}{ccccccccc}
\hline \hline
{Flaring star} & {Ecl. star} & $\theta$($\degr$) &$\lambda$($\degr$)  & 
$h$ (R$_\odot$) & $h-R_*$ (R$_*$) & $R_3$ (R$_\odot$) & $\log n_{\rm e}$ (cm$^{-3}$) & $B$ (G)\\
\hline
Sec & Pri & $+12.5 - +14.3$ & 173.3--176.2 & 1.53--1.66 & 0.65--0.78 &
0.002--0.012 & 12.8--14.0 & 800--3300\\
Sec & Sec & $-17.5 - +49.0$ & 20.8--25.2  & 0.95--1.87 & 0.02--1.01 &
0.010--0.273 & 10.7--12.9 & 76--910 \\
``  & ``  & $-49.9 - -39.1$ & 20.8--25.2  & 1.63--1.87 & 0.75--1.01 &
0.084--0.223 & 10.9--11.5 & 88--180 \\
Sec & S+P & $-52.7 - -42.9$ & 20.8--25.2  & 1.62--1.87 & 0.74--1.01 &
0.084--0.197 & 11.0--11.5 & 98--180 \\
Pri  & Pri & $+37.5 - +55.0$ & 20.7--25.2  & 0.50--0.54 & 0.00--0.08 &
0.002--0.014 & 12.7--14.0 & 710--3000 \\
\multicolumn{2}{c}{All~results} & $-53 - +55$ & 20.7--25.2 or 175
& 0.50--1.87 & 0.00--1.01 & 0.002--0.273  & 10.7--14.0 & 76--3300\\
\hline
\end{tabular}
\end{footnotesize}
\end{center}
\end{table*}

\begin{figure*}
   \centering
   \includegraphics[width=0.90\textwidth]{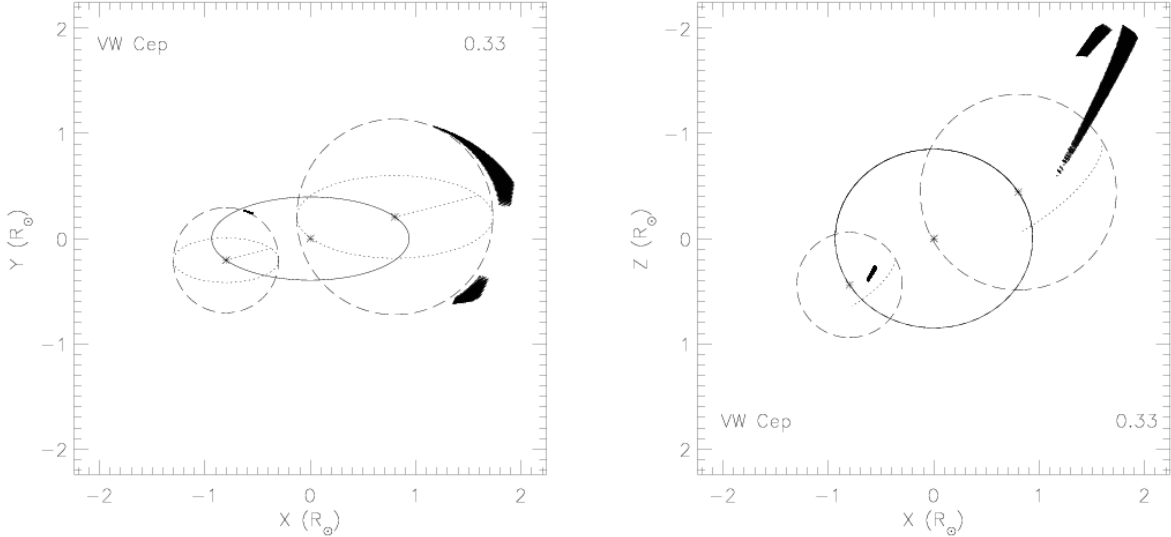}
   \caption{Relative positions of the VW~Cep stars (dashed lines) during
   first contact ($\phi$=0.337), as seen from the front (left
   panel) and top 
   (right panel). All the possible results are represented
   with black points indicating the {\em center} of the emitting
   region. A dotted line marks the equator (left panel) and the
   meridian origin of longitude (both panels). Longitude goes positive
   counter-clockwise.}
   \label{vwcepsols}
\end{figure*}

At the first and fourth contacts the distance between the center
of the flare and the center of secondary star must be:

\begin{displaymath}
\Delta_1=\sqrt{(R_2+R_3)^2-(y_3-y_2)^2}
\end{displaymath}
and similarly, at the second and third contact:
\begin{displaymath}
\Delta_2=\sqrt{(R_2-R_3)^2-(y_3-y_2)^2}
\end{displaymath}

\begin{figure*}
   \centering
   \includegraphics[width=0.90\textwidth]{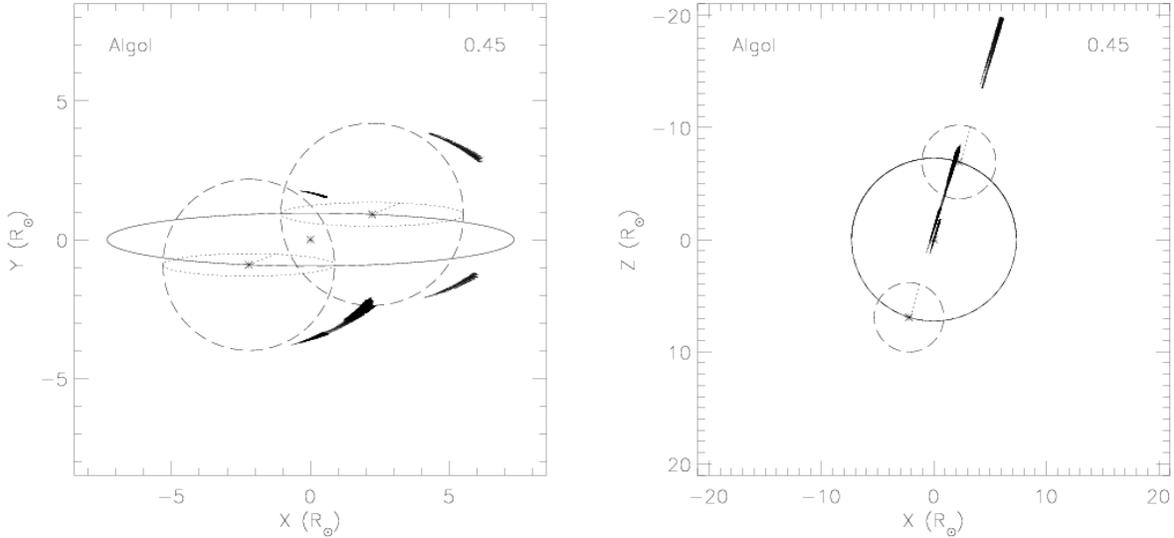}
   \caption{Relative positions of the Algol stars (dashed lines) during
   first contact ($\phi$=0.451) in the Beppo/SAX observations, as seen
   from the front (left panel) and top 
   (right panel). All the possible results are represented
   with black points indicating the {\em center} of the emitting region.}
   \label{algol1sols}
\end{figure*}

Thus, we will have four equations: $x_3=x_2+\Delta_1,
\quad  x_3=x_2+\Delta_2, \quad x_3=x_2-\Delta_2, \quad x_3=x_2-\Delta_1$
that we can use to calculate numerically 
the times of the four contacts for each point of the grid.
Similar equations can be easily derived for the other configurations
of the flare. 
Some geometrical constraints can be added to ensure the totality of
the eclipse ($y_3 + R_3 \le y_2 + R_2$) and a valid position in $z$
($z_3 +\Delta_1 < z_2$). 
In cases with large values of $h$ we have imposed another constraint: 
the centrifugal
force must be lower than the resulting gravitational forces of the two
stars. In this way, 
$h + R_3$ must be lower than a certain value that is very close to
$a$. 
Further restrictions can be imposed regarding the light curve observed
during the phases in which no eclipses where observed: 
$(x_3-x_2)^2 + (y_3-y_2)^2 > R_2+R_3$
(substitute the subindex $2$ with $1$ to avoid eclipses by the primary). 
The results found are shown in Tables~\ref{tab:vwcep}, \ref{tab:algol}
and in Figs.~\ref{vwcepsols}--\ref{algol2sols}. More detailed results
are found in Figs.~\ref{vwcepres}-\ref{algolresbis}, and
Fig.~\ref{simulations} show several animations of valid solutions in
the three flares.  

The physical interpretation of an emitting region shaped as a sphere
detached from the surface could be the apex of the loops involved in
the flare, with the emitting region being the place of magnetic
reconnection. Although the real shape is unlikely to be a sphere,
such a form is the simplest approximation and would be the smallest sphere
that contains the real emitting region. We have also conducted a test
using the shape of a loop (see Sect.~\ref{sect:loop} and
Sect.~\ref{sect:discussion}), and another test considering only the
central points of the eclipse (See Sect.~\ref{sect:discussion}), yielding
similar results. We therefore conclude that the use of a sphere gives
valid results. 

An additional result can be obtained from the temperature ($T$) and emission
measure ({\em EM}) of the flare.  
We can calculate the electron density ($n_{\rm e}$) of
the emitting region by calculating the {\it EM}  during the
flare. In this case we use the values of the $T$ and {\it EM}
determined in the original papers when available. 
Since {\it EM}$\sim 0.8 \, n_{\rm e}^2 \, V$, it is 
possible to get the density $n_{\rm e}$ from the net flare {\it EM}, 
using $V=4/3 \pi R_3^3$. If we consider that the magnetic pressure, 
$B^2/(8 \pi)$, should be at least as large as the electron pressure
($2 n_e k T$), we can get the minimum magnetic field needed
to have a stable structure in the flare. In the cases in which the
whole sphere of the flaring region is not observable, we have
corrected the volume $V$ by a factor accounting for the visible
fraction. The results of the $n_{\rm e}$ and $B$ are listed in
Tables~\ref{tab:vwcep}--\ref{tab:algol}.
Next we describe the particular details of each of the three eclipses.

\begin{table*}
\caption{Algol (B8V/K2III, $R_{1,2}$=3.09/3.29~R$_\odot$) range of possible
solutions of the eclipses of the flares in 1997 and 2002.}\label{tab:algol} 
\begin{center}
\begin{footnotesize}
 \begin{tabular}{cccccccc}
\hline \hline
{Observation} & $\theta$($\degr$) &$\lambda$($\degr$)  & 
$h$ (R$_\odot$) & $h-R_2$ (R$_*$) & $R_3$ (R$_\odot$) & $\log n_{\rm e}$ (cm$^{-3}$) & $B$ (G)\\
\hline
BeppoSAX & $ +12.7 - +13.9 $ & 178.6--179.8  & 7.81--12.5 & 1.37--2.80 & 0.033--0.33
& 11.8--13.3 & 630--3500 \\
   ``    & $ -87.98 - -22.5 $ & 10.10--179.2  & 3.14--9.45 & $-0.05$\,--\,+1.87 & 0.26--1.69
& 10.8--11.9 & 220--750 \\
BeppoSAX (self-eclipsed) & $+1.0 - +15.6$ & 0.4--1.6 & 7.50--13.60 &
1.28--3.13 & 0.29--1.19 & 10.9--11.8 & 240--690 \\
 ``                      & $-30.0 - -16.4$ & 0.4--1.6 & 7.75--13.60 &
1.36--3.13 & 0.30--1.13 & 11.0--11.80 & 250--670 \\
XMM-Newton & $ +6.9 - +16.0$ &  197--272 & 3.75--7.31 & 0.14--1.22 & 0.078--0.31
& 11.0--11.9 & 150--420 \\
 ``        & $-50.5 - -24.8$ &  198--323 & 5.82--10.7 & 0.77--2.25 & 0.093--0.76
& 10.4--11.8 & 76--370 \\
\hline
\end{tabular}
\end{footnotesize}
\end{center}
\end{table*}

\subsection{VW Cep}\label{sec:vwcep}
The eclipse detected in VW Cep with ASCA was analyzed by
\citet{cho98}. They concluded that the dip observed in the light
curve (see Fig.~\ref{vwceplc}) corresponds to an eclipse of the
flaring region. Since
the decay of the flare lasts almost one orbital period, they assume
that the flare must occur in a polar region because otherwise it would
be occulted by one of the stars out of the eclipse. Therefore
\citet{cho98} conclude that the flare occur in a polar region of the
primary. They calculate also the size scale of the flare from the
ingress and egress times yielding a size of $\sim 5.5\times
10^{10}$\,cm ($\sim 0.8 R_\odot$), ignoring the rotational velocity
of the secondary star because the flare is assumed to be polar. 
Rotational modulation will however be produced only if
the emitting region of the flare is close enough to the stellar
photosphere to be occulted. 

We have reanalyzed the data, and determined that the phases of the
four contacts are $\phi$=0.337, 0.374, 0.486 and 0.548   
respectively, using the elements
provided by \citet{alu94}, where $P_{\rm orb}$=0.2783076 and
$T_0$=2,448,862.5220 (at $\phi$=0 the G5V star is behind the K0V star),
and the stellar data given by \citet{hil89}.
If we assume that
the dip observed corresponds to an eclipse (rather than the decay of a
flare followed by a second flare), then this eclipse partially
overlaps with the secondary optical eclipse, and therefore it is likely that
a flare in the secondary is eclipsed by the primary. We have
considered also the cases in which the flare is self-eclipsed by the
host star. If the emitting region is a flare in the primary
separated from the 
stellar surface, it is not possible to distinguish it from a flare in
the secondary occulted by the primary. The results of the
analysis (displayed
in Fig.~\ref{vwcepsols} and Table~\ref{tab:vwcep}) indicate that no
flare in polar regions ($|\theta|>55\degr$) can produce the observed
light curve, and that the 
flare could be either associated with the primary or the secondary. Besides,
the size of the emitting region is always $R_3<0.3$\,R$_\odot$, well
below the calculations of \citet{cho98}. 

We have calculated the electron density based on the {\em EM} during
the flare. We have used the same values calculated by
\citet{cho98}: 
$\log T_{1,2}=6.87, 7.35$ (K) and $\log EM_{1,2}=52.56,
52.58$ (cm$^{-3}$). Results are displayed in Table~\ref{tab:vwcep}.

Although the SIS and GIS light curves are
suggestive of an eclipse of a flare, we cannot reject the possibility
that there are actually two flares. However we believe that the slight
increase in the slope at the ``first contact'' might be an
indication of an eclipse rather than a fast decay of a first flare
\citep[the SIS light curve seemed to indicate more clearly the eclipse
interpretation, see][]{cho98}.

\subsection{Algol with BeppoSAX}
BeppoSaX detected a very remarkable flare in Algol in 1997, with an
increase in X-ray flux by 2 orders of magnitude. Besides,
the flare was eclipsed during the decay in a rather clean light
curve. 
\citet{schm99} analyzed these data, and 
they concluded that the flaring region was placed near the
southern pole. 
If we apply our method to the same data \citep[using the same phases
calculated by][]{schm99}, we have a set of results in a wide range of
latitudes, and even some at $\theta\sim 15.6\degr$ in the northern
hemisphere. Among the results found, those in the northern hemisphere
with the flare eclipsed by the primary
imply a relatively large distance from the secondary star ($h$), and
such positions are likely hampered by gravitational forces 
of the secondary star and/or some mass transfer given the low latitude
and short distance of these solutions to the center of mass of the
system. 
However we did not compute the effects of gravity
in order to reject these solutions. 
Another set of results in the southern hemisphere indicates that the
flaring region could still be present at latitudes between $-22$ and
$-59\degr$, in this case at lower heights over the secondary, and likely
less affected by mass transfer processes. 
Finally, those results related to a self-eclipse by the secondary
star imply large values of $h$ and low latitudes, but no mass transfer
can be involved in this case (see Table~\ref{tab:algol}).

\begin{figure*}
   \centering
   \includegraphics[width=0.90\textwidth]{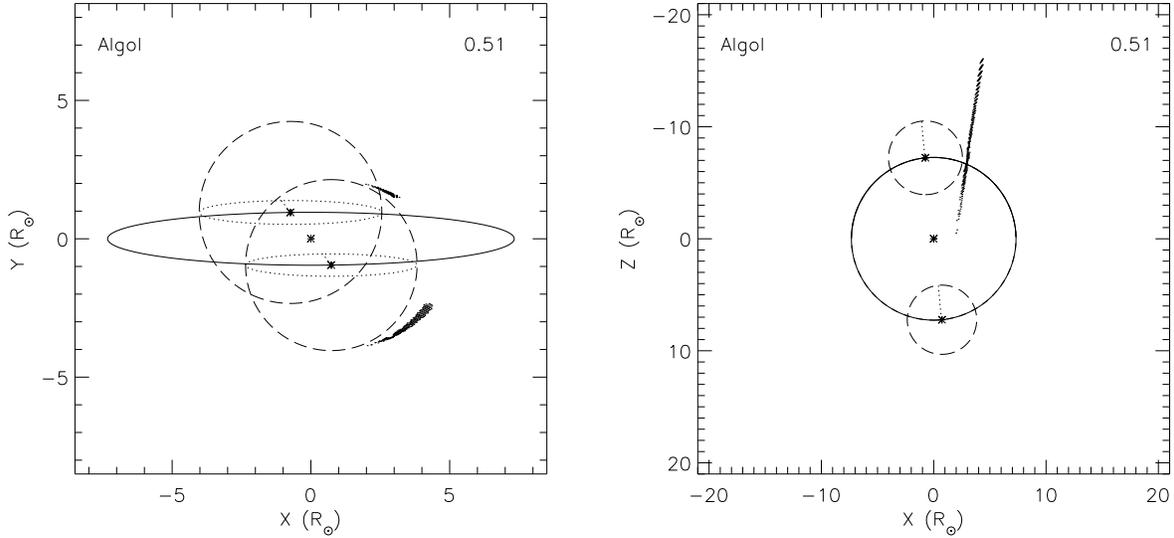}
   \caption{Relative positions of the Algol stars (dashed lines) during
   first contact ($\phi$=0.516) in the XMM observations, as seen from the front (left 
   panel) and top 
   (right panel). All the possible results are represented
   with black points indicating the {\em center} of the emitting region.}
   \label{algol2sols}
\end{figure*}

The phases of the eclipse (0.451, 0.488, 0.505, 0.545)
make this analysis more complicated than the others. While in the other
cases the range of solutions does not depend on the approximation of a
spheric shape for the emitting region, in this particular case
(totality happens while there is a stellar conjunction) it is
possible to find other solutions
consisting in an eclipse of a flare that is very
close to the surface. In order to model this case we assumed that a
sphere partially below the photosphere of the star might 
produce the same light
curve, and therefore we have relaxed the conditions to allow this
geometry to happen. This workaround allows us the use of
spheric caps in the analysis as well.  This situation is not considered
for the other two eclipses (VW Cep and Algol with {\em XMM-Newton}) because the
geometry of the eclipses is not compatible with this configuration: the Algol case is
explained below, and the second contact of the coronal eclipse of
VW~Cep occurs when the photospheric eclipse
starts (see Fig.~\ref{algol1lc}), and therefore only a region detached from
the surface can be totally eclipsed at that time. 
The electron density under this configuration 
has been calculated using
the volume of the emitting region that is visible (note that in some
cases the emitting region is partially occulted below the secondary
star during the whole flare). 
The solutions found under this configuration show that the flare could
be present at almost any longitude near the south pole of the
star, producing the same kind of light curve eclipse as the other
results mentioned in former paragraph.

\subsection{Algol with XMM-Newton}
{\em XMM-Newton} detected a flare in Algol in 2002 that was analyzed by
\citet{sch03}, who concluded that an eclipse of the flare was taking
place, indicating that the flare should have occurred near the limb in
the northern hemisphere since the first contact takes place at
$\phi >$0.5. In this case \citet{sch03} used a reconstruction of the
flaring plasma region using the whole light curve, concluding that the
flare takes place at a 
height of $\sim 0.1 R_*$ ($R_2=3.29 R_\odot$) and
$\theta < 69\degr$. 

We have determined the phases of the four contacts at 0.516, 0.527,
0.554, 0.565 using the ephemeris of \citet{aln85}, where $P_{\rm
  orb}$=2.8673285~d and $T_0$=2,445,739.003. These values are in
agreement with those proposed by 
\citet{sch03}, although they only expressed the first three contacts
and assumed that the fourth contact takes place at the end of the
observation. We do not expect this to affect the range of solutions.
However there is an alternative
interpretation of the light curve: the eclipse could actually be
starting at phase 0.506 (the apparent peak of the flare), while the
change in slope at $\phi \sim 0.516$ could actually correspond to the flare
maximum emission. The determination of the fourth contact has also some
uncertainty. The analysis of the light curves of different energy
bands \citep{sch03} seems to suggest that the eclipse following a
flare maximum is 
the best option, so we will refer to this situation from now on. 

Given the phases involved in the eclipse (see
Fig.~\ref{algol2lc}) it is easier to analyze this flare than the BeppoSAX
flare, since a configuration of an emitting region overlapping with the
star is not possible: the eclipse starts after $\phi\sim 0.5$,
invalidating the solutions in the southern hemisphere, and those
in the northern hemisphere were rejected in the analysis.
We only found valid solutions for the configuration in which the
flare (hosted by the secondary) is eclipsed by the primary
star. Solutions, including the result reported by \citet{sch03}, are
possible both in the northern ($7\degr \la 
\theta \la 16\degr$) and southern hemisphere ($-50\degr \la \theta \la
-25\degr$), 
and given the position of the flare (Figs.~\ref{algol2sols},
\ref{algolresbis}) in most cases it is unlikely that the flaring
region is affected by the mass transfer between the two stars.

Since \citet{sch03} do not report the value of the {\it EM},
we made a spectral fit to the spectrum in the quiescent emission
(log~{\it EM}~[cm$^{-3}$]=53.64) and 
in the flaring non-eclipsed emission (log~{\it EM}~[cm$^{-3}$]=53.89),
resulting in a net value for the flare of log~{\it EM}~[cm$^{-3}$]=53.53 at
log~$T$~[K]$=7.50$. The range of possible values of the electron
density (Table~\ref{tab:algol}) are consistent with those reported by
\citet{sch03}, but unlike \citet{sch03} we also found solutions in the
southern hemisphere, and a wider range of solutions in the northern
hemisphere. 
 
\onlfig{8}{
\begin{figure*}
   \centering
   \includegraphics[width=0.95\textwidth]{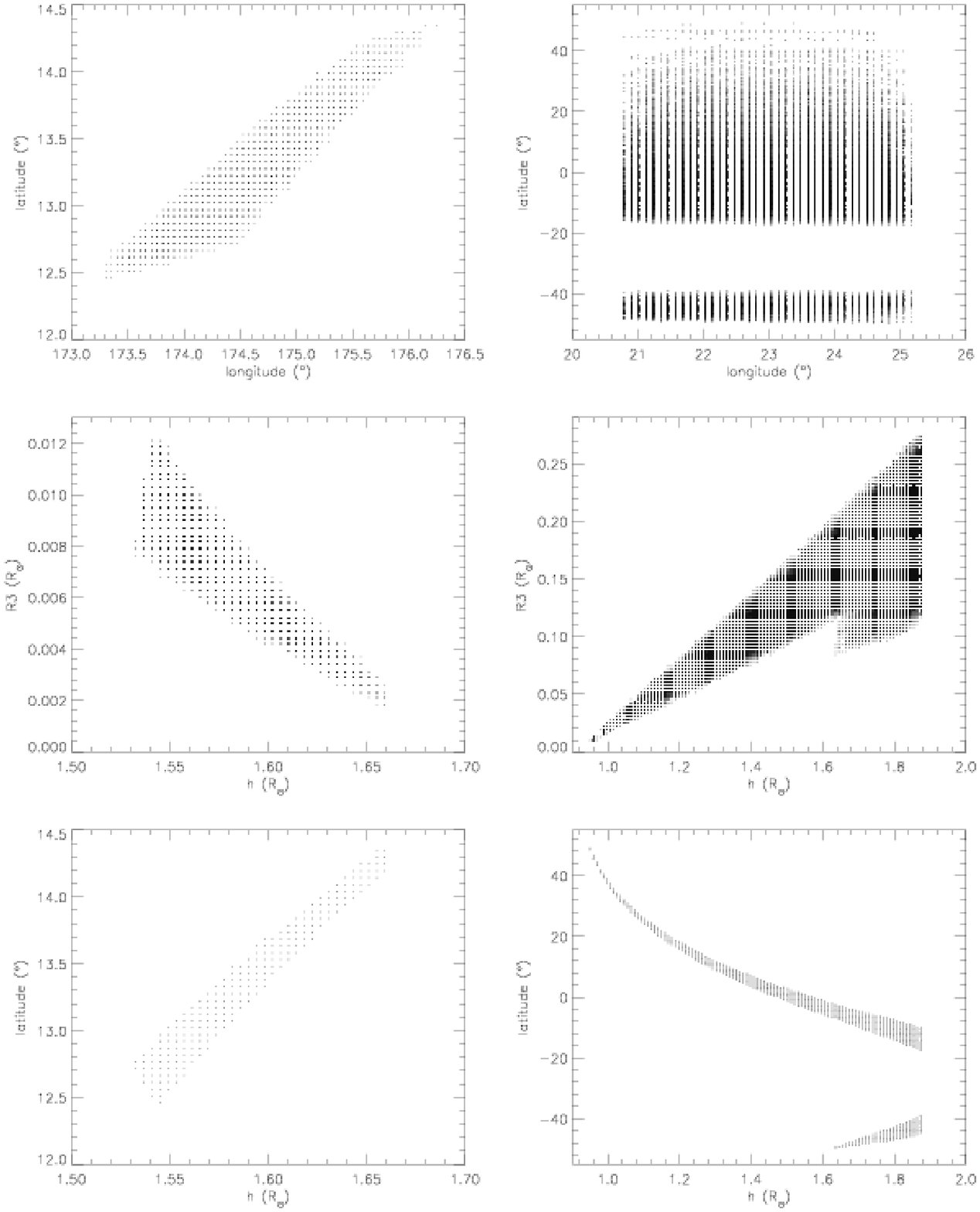}
   \caption{Results of the flare in VW Cep, flare in the
     secondary. {\em Left:} eclipsed by primary. {\em Right:}
     self-eclipsed by secondary.}
   \label{vwcepres}
\end{figure*}
}

\onlfig{9}{
\begin{figure*}
   \centering
   \includegraphics[width=0.95\textwidth]{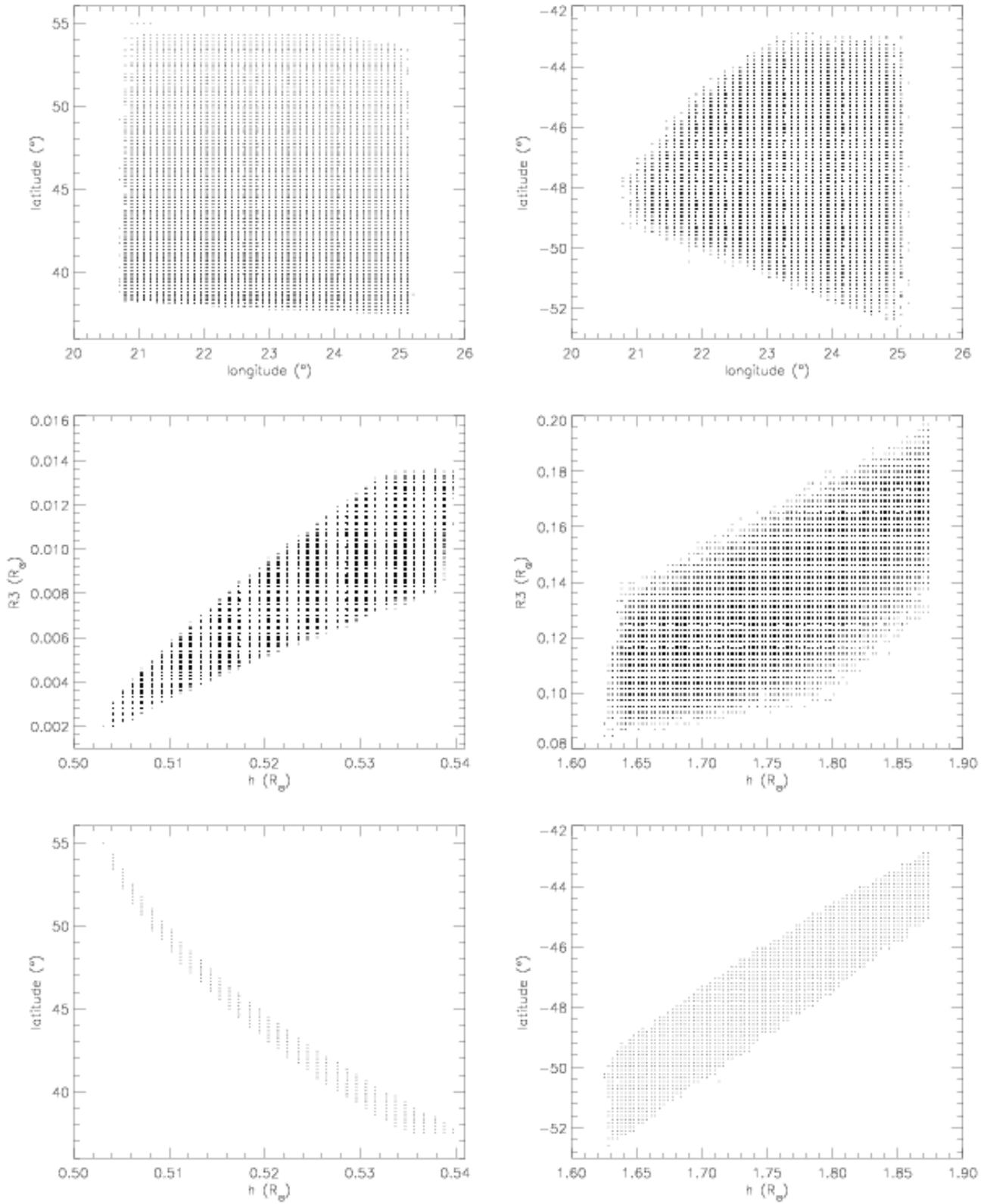}
   \caption{Results of the flare in VW Cep. {\em Left:} flare in
     primary, self-eclipsed by primary. {\em Right:} flare in
     secondary, the eclipse starts behind secondary and ends behind
     the primary.}
   \label{vwcepresbis}
\end{figure*}
}

\onlfig{10}{
\begin{figure*}
   \centering
   \includegraphics[width=0.95\textwidth]{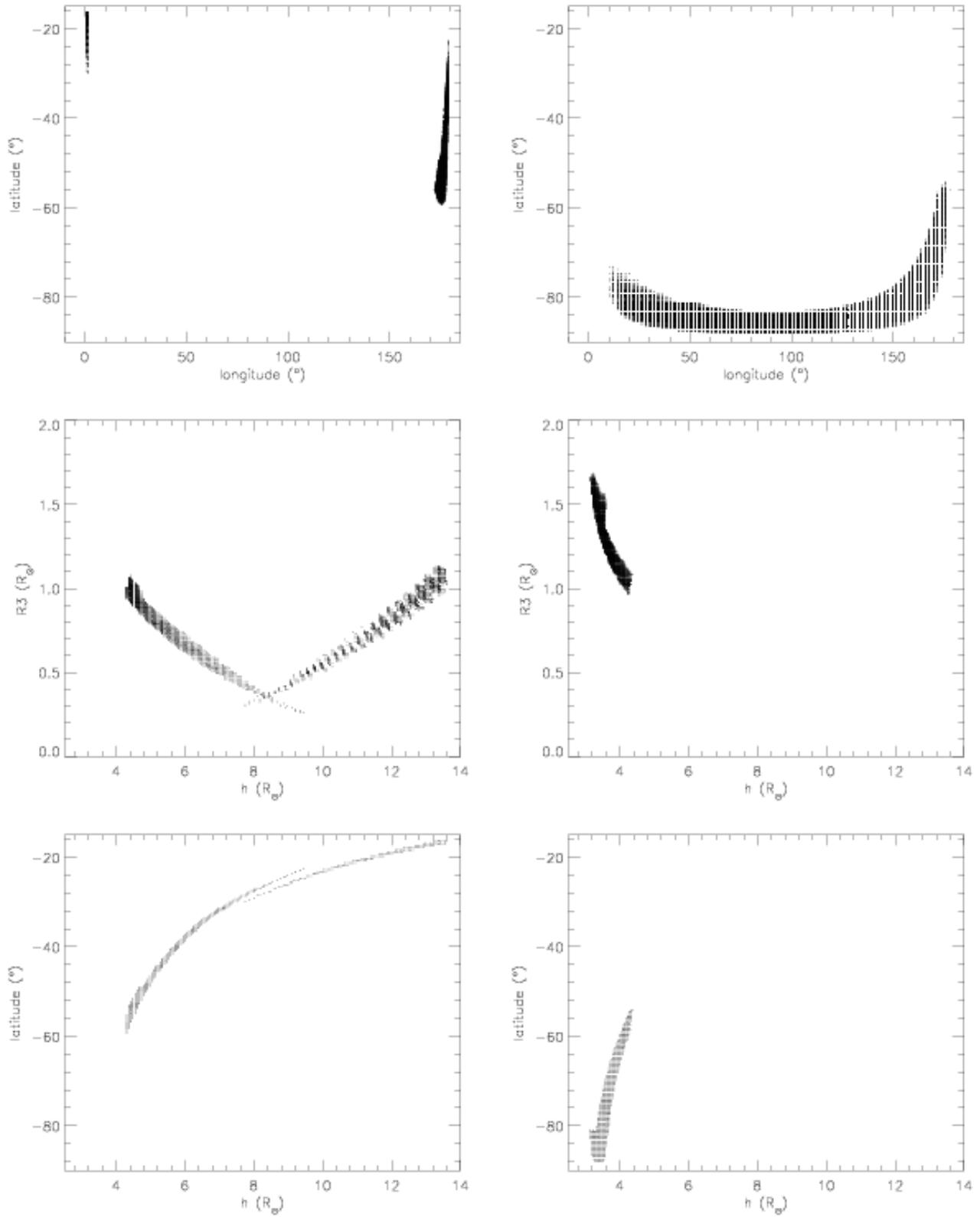}
   \caption{Results of the flare in Algol observed with BeppoSAX,
     solutions in the southern hemisphere. {\em Left:} The emitting
     region is detached from the surface of the star (the results
     around $\lambda \sim 1\degr$ correspond to self-eclipse by the
     secondary). {\em Right:} The 
     emitting region can overlap with the star (therefore it can be a
     spheric cap).} 
   \label{algolres}
\end{figure*}
}

\onlfig{11}{
\begin{figure*}
   \centering
   \includegraphics[width=0.45\textwidth]{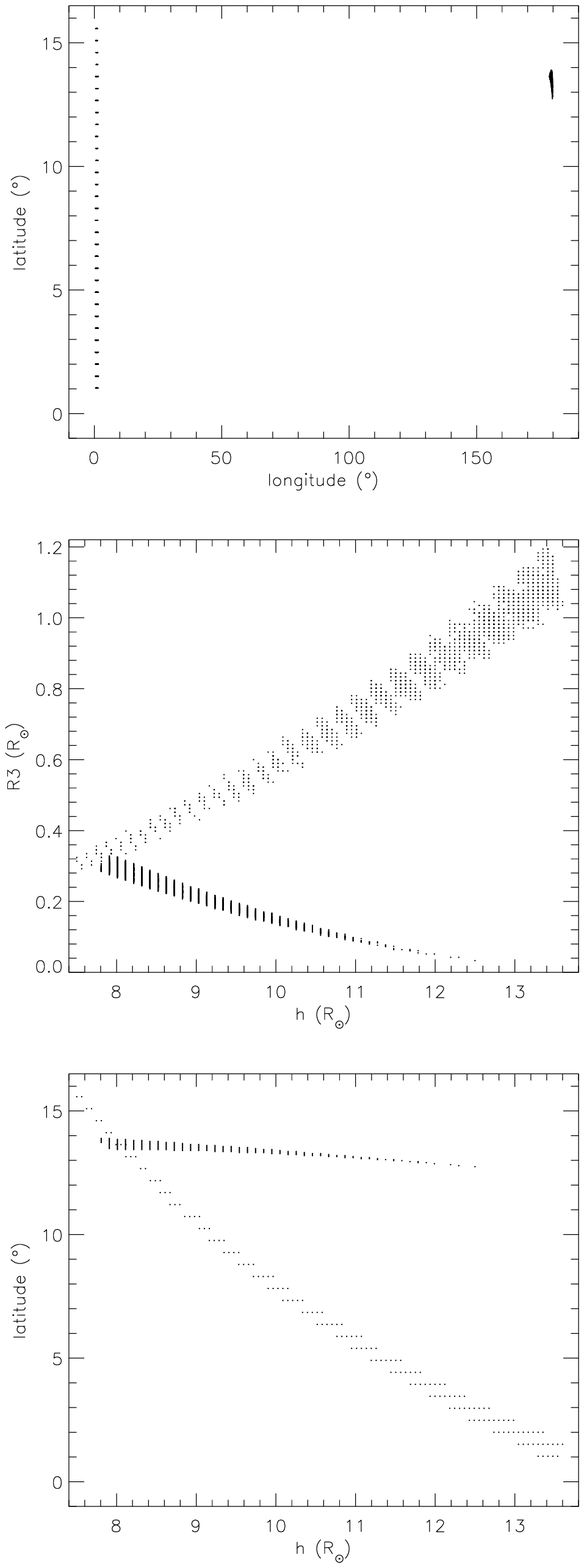}
   \includegraphics[width=0.45\textwidth]{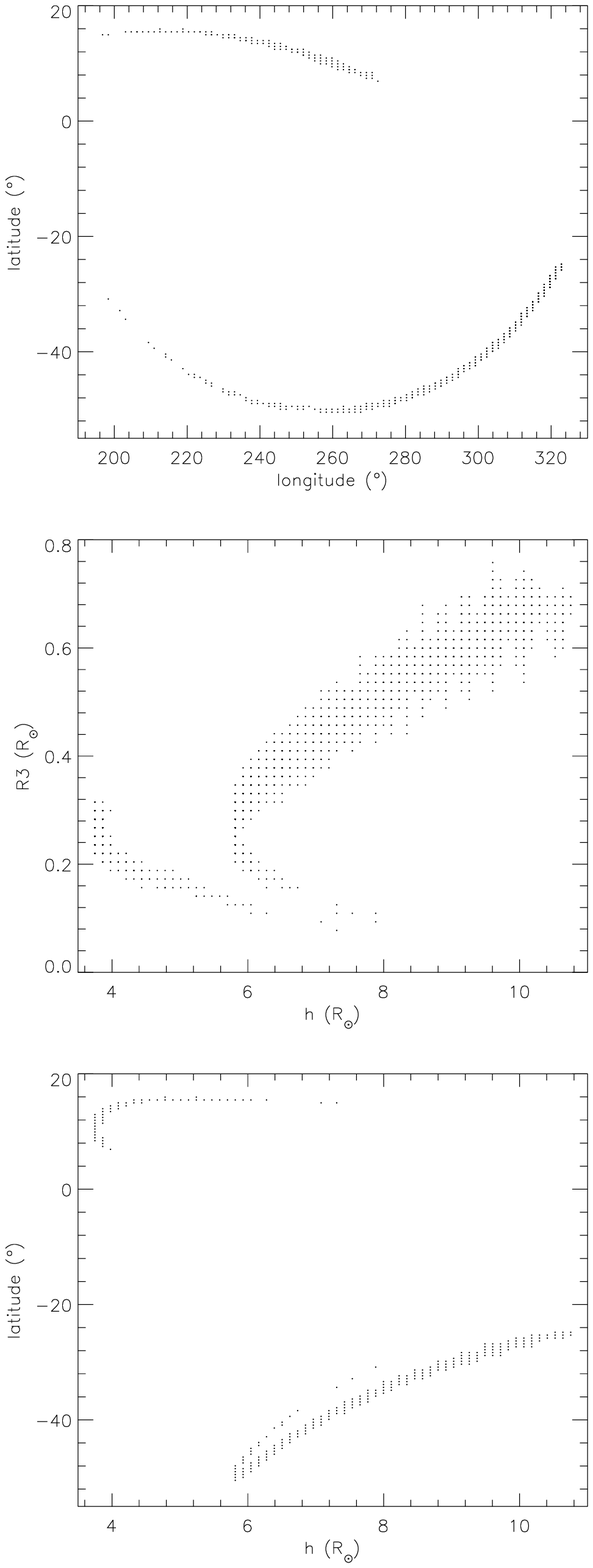}
   \caption{Results of the flare in Algol. {\em Left:} Observed with
     BeppoSAX, solutions in the northern hemisphere (the results
     around $\lambda \sim 1\degr$ correspond to self-eclipse by the
     secondary). {\em Right:}  Observed with {\em XMM-Newton.}}
   \label{algolresbis}
\end{figure*}
}

\onlfig{12}{
\begin{figure*}
   \centering
   \caption{Several examples of animations of valid results with
     different configurations. A white circle (or loop) indicates the
     emitting region of the flare being eclipsed. The latitude and
     longitude used in the simulations are 
     shown in the upper panel, while the light curve is displayed
     in the lower panel.}
   \label{simulations}
\end{figure*}
}

\begin{figure*}
   \centering
   \includegraphics[width=0.85\textwidth]{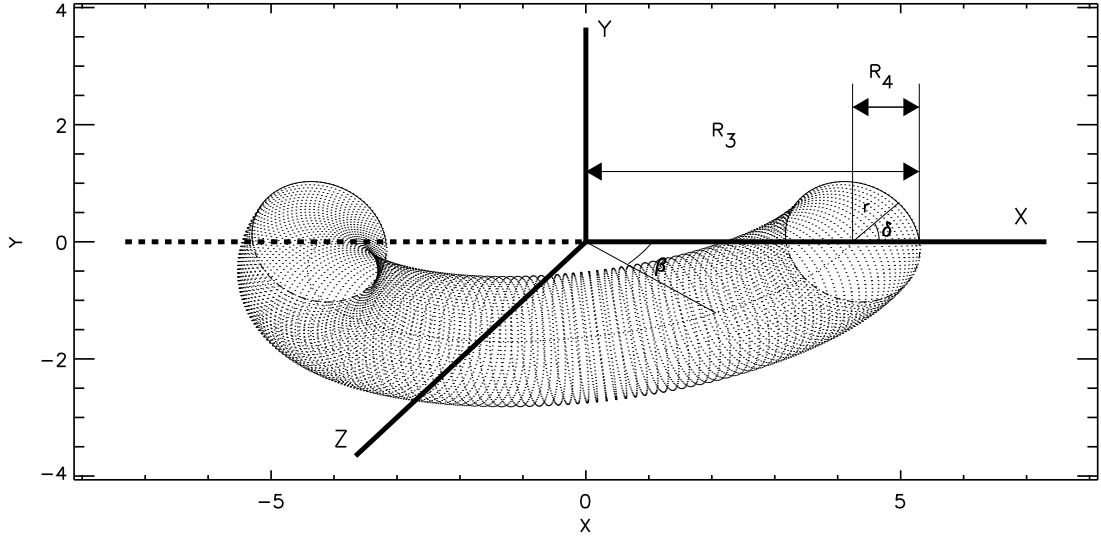}
   \caption{Schematic view of the coronal loop. The loop lies around the
     XZ plane. The Z axis points
     opposite from the center of the host star and Y axis is paralel to the
     closest meridian.}
   \label{fig:geometryarc}
\end{figure*}

\subsection{Eclipses of coronal loops}
\label{sect:loop}
The shape chosen for the analysis has been the simplest possible, a
sphere. The use of a different geometry will further restrict the
number of solutions. Since there is an infinite range of shapes
possible, we have chosen only one particular case to show the deviation
from the solutions found with the sphere. We chose a shape similar to
what is observed in solar coronal loops. Following the notation used
in the paper we have defined a torus contained in a sphere of radius
$R_3$ (the external radius of the torus), a cross-section radius fixed
to $R_4 = 0.2\, R_3$, and a circular curvature. 
The arc is placed perpendicular to
the stellar surface (see Fig.~\ref{fig:geometryarc}), with a local
reference frame with the Z axis
pointing opposite from the center of the star and the Y axis parallel
to the tangent of the nearest meridian. In this local reference frame we
can define each of the points ($x_{\rm i}, y_{\rm i}, z_{\rm i}$) of 
the loop using two angles ($\beta$, $\delta$) and a radial variable
($r$, the radius of the cross-section) as follows:
{\setlength\arraycolsep{1pt}
\begin{eqnarray*}
x_{\rm i} & = & (R_3 - R_4 + r \cos \, \delta) \cos \, \beta , \quad
y_{\rm i} = r \sin \, \delta, \\
z_{\rm i} & = & (R_3 - R_4 + r \cos \, \delta) \sin \, \beta
\end{eqnarray*}}
where r varies from 0 to $R_4$, $\beta$ from 0$\degr$ to 180$\degr$
and $\delta$ from 0$\degr$ to 360$\degr$.
In order to account for the eclipses coverage we need to transform the
local coordinates ($x_{\rm i}, y_{\rm i}, z_{\rm i}$) into the
coordinates of the rest reference frame used for the stellar system
(Fig.~\ref{fig:eclipse}). We need to apply 3 rotations, in latitude
($\theta$), in longitude ($\lambda + \phi$), and in inclination ($i$):

a rotation in latitude (around X),
{\setlength\arraycolsep{1pt}
\begin{eqnarray*}
x' = x, \quad
y' = (\cos \, \theta) \, y + (\sin \, \theta) \, z, \quad
z' = (-\sin \, \theta) \, y + (\cos \, \theta) \, z,
\end{eqnarray*}}
a rotation in longitude (around Y),
{\setlength\arraycolsep{1pt}
\begin{eqnarray*}
x'' & = & \cos (\phi + \lambda) \, x' + \sin (\phi + \lambda) \, z', \quad
y'' = \, y', \\
z'' & = & -\sin (\phi + \lambda) \, x' + \cos (\phi + \lambda) \, z',
\end{eqnarray*}}
and a rotation due to inclination (around X),
{\setlength\arraycolsep{1pt}
\begin{eqnarray*}
x''' & = & x'', \quad
y''' = (\sin \, i) \, y'' - (\cos \, i) \, z'', \\
z''' & = & (\cos \, i) \, y'' + (\sin \, i) \, z'',
\end{eqnarray*}}

Therefore, the position of a point of the loop 
in the same reference frame used in the problem will be: 
{\setlength\arraycolsep{1pt}
\begin{eqnarray*}
x'''_{\rm i} & = & x_3 + \cos (\phi + \lambda) \, x_{\rm i} + \sin
(\phi + \lambda) \, (-\sin \, \theta \, y_{\rm i} + \cos \, \theta \,
z_{\rm i}), \\ 
y'''_{\rm i} & = & y_3 + \sin \, i \, (\cos \, \theta \, y_{\rm i} + \sin \,
\theta \, z_{\rm i}) - \cos \, i \, (-\sin (\phi + \lambda) \,
x_{\rm i} + \\ & &
\cos (\phi + \lambda) \, (-\sin \, \theta \, y_{\rm i} + \cos \,
\theta \, z_{\rm i}), \\ 
z'''_{\rm i} & = & z_3 + \cos \, i \, (\cos \, \theta \, y_{\rm i} + \sin \,
\theta \, z_{\rm i}) + \sin \, i \, (-\sin (\phi + \lambda) \,
x_{\rm i} + \\ & &
\cos (\phi + \lambda) \, (-\sin \, \theta \, y_{\rm i} +
\cos \, \theta \, z_{\rm i}),
\end{eqnarray*}}

\begin{figure*}
   \centering
   \includegraphics[width=0.90\textwidth]{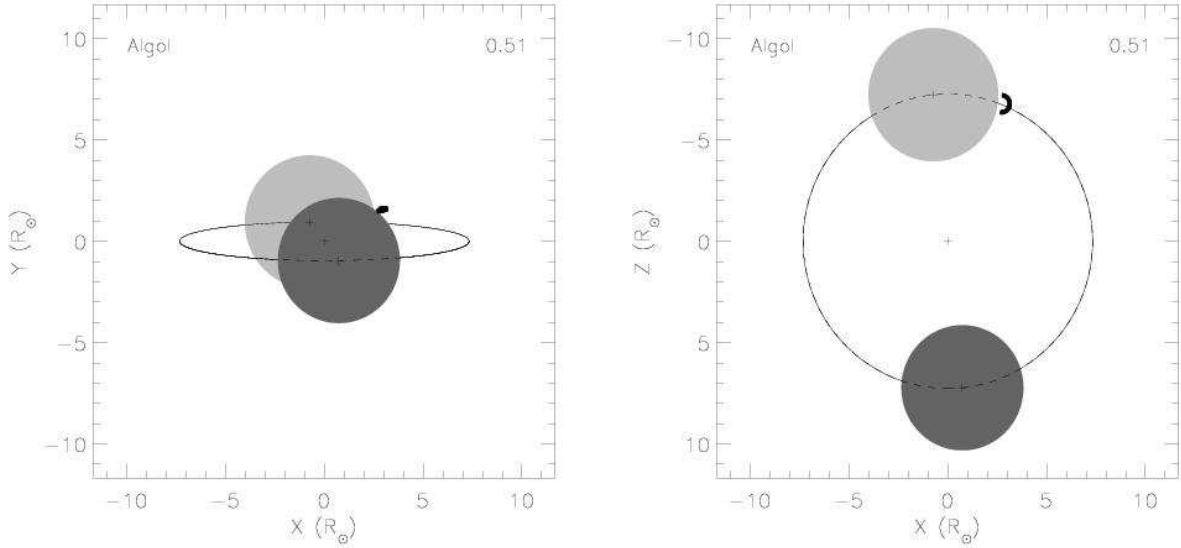}
   \caption{An example of a valid solution using a coronal loop placed
   at $\theta =10.1 \degr$, $\lambda = 258.2 \degr$, $R_3=0.53\,{\rm
     R}_\odot$ and $h =3.52\,{\rm R}_\odot$. Views from the front
   (left panel) and the top (right panel) during first contact.}
   \label{fig:examplearc}
\end{figure*}

We have applied this method to search for the results in the case of
the flare observed with {\em XMM-Newton} in Algol.
In Fig.~\ref{fig:examplearc} we display a valid solution near the
equator, with a simulation included in Fig.~\ref{simulations}. The
possible solutions (Fig.~\ref{fig:examplearcresults}) are very similar
to those found for a spheric shape 
(the $h$ and $R_3$ are slightly different since $R_3$ has a
different definition). The density derived from this geometry would
also be higher than that derived from the spheric case.

\onlfig{15}{
\begin{figure*}
   \centering
   \includegraphics[angle=90,width=0.90\textwidth]{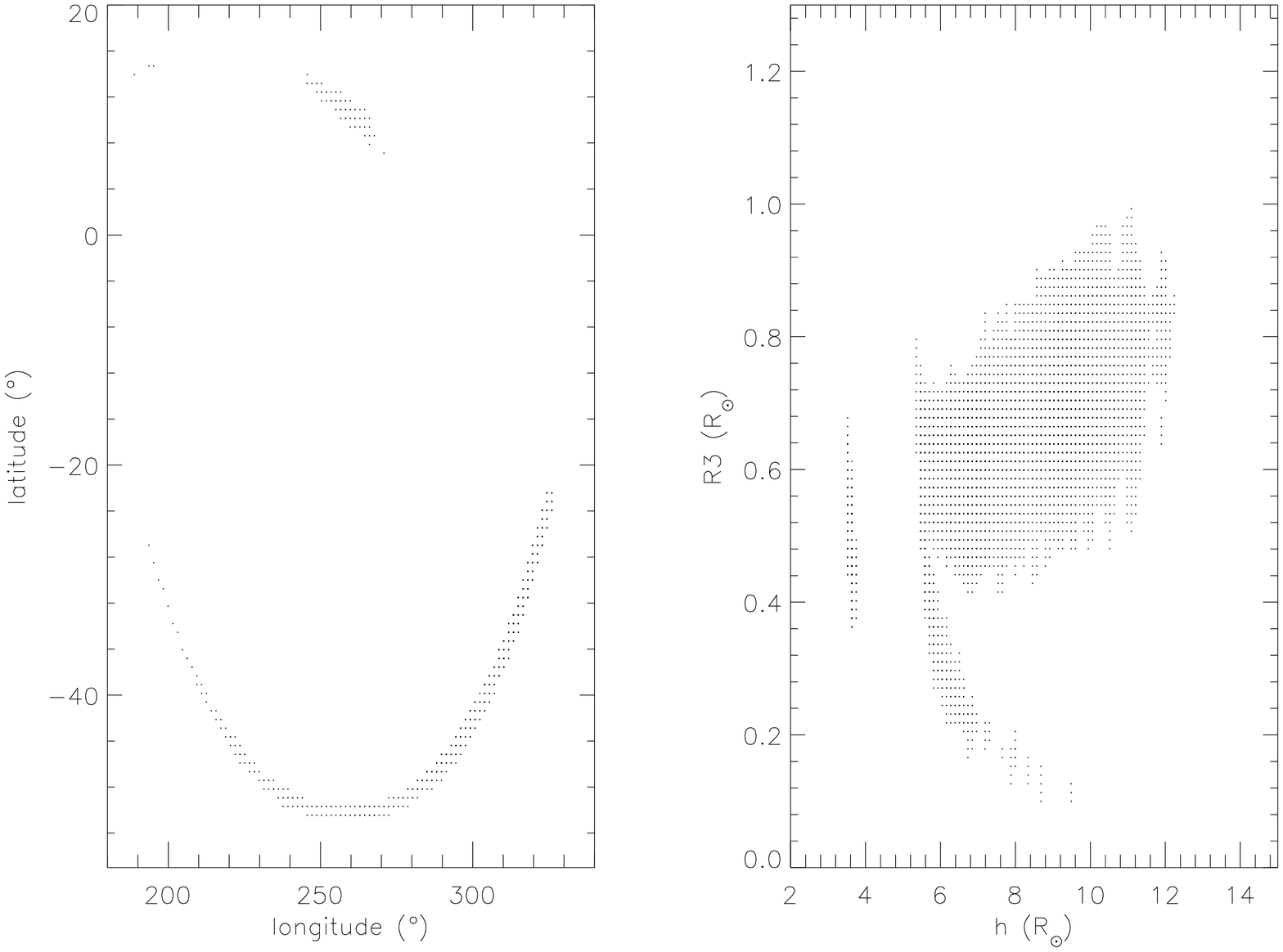}
   \caption{Range of valid solutions for the case of a coronal
     loop in the {\em XMM-Newton} flare with the proposed orentation.
     Flare happens in secondary, and it is eclipsed by primary.}
   \label{fig:examplearcresults}
\end{figure*}
}

\section{Discussion}
\label{sect:discussion}
The approach presented here constrains the geometry (location and size) of the flaring region for eclipsed flares. A
simple calculation also yields a lower limit of the electron
density (the plasma does not need to be homogeneously distributed in a
sphere). The results are consistent with the range of values derived from
density-sensitive line ratios in active stars. 
The method assumes that the emission measure remains constant during
the eclipse (this is important only for the calculation of electron
density and magnetic field), and that the
spherical shape of the emitting region remains unchanged during the eclipse. 
This shape has been chosen as the simplest possible approximating a
compact region.
Spheres with a filling factor lower than 1 would imply higher values
of the electron density, and we do not expect that
other shapes vary substantially the results of latitude, longitude
and height, except in cases like the {\em BeppoSAX} Algol flare, where the
flaring region can be actually attached to the surface and have a
significant surface filling factor. 
The solutions where the emitting region is detached from the surface
  could correspond to the apex of the 
loops involved in the flare, although magnetic reconnection in the
solar flares does not happen necessarily at the apex, and the site of
reconnection does not need to be the site of maximum X-ray emissivity.
Although it is unlikely that the emitting region is a sphere, we
consider this a good approximation in order to search for the
location. 
We have conducted two tests to quantify the
relevance of the shape chosen with respect to the real solutions. 
A first test was conducted using a coronal loop instead of a sphere
(see Sect.~\ref{sect:loop}),
and considering the four contacts of the eclipse. The orientation of
the loop (with the shape of a torus) was fixed, as well as the
cross-section radius (0.2\, $R_3$). The set of results is similar to
that of spherical shape, but the range of solutions is smaller.
In the second test, we 
have considered the simpler case in which only two contacts are used
in the model. These contacts would correspond to the ingress (midpoint
between $\phi _1$ and $\phi _2$) and egress (midpoint
between $\phi _3$ and $\phi _4$) of the center of the emitting region,
and therefore independent of its shape. The test conducted
has shown that the range of solutions is similar to
those obtained for the restrictions imposed by
the use of four contacts and a spherical shape. The results are displayed in
Figs.~\ref{fig:test1}--\ref{fig:test3}.
These two
tests confirms that the approximation of a sphere can be valid
also as an approximation for more complicated shapes.

\onlfig{16}{
\begin{figure*}
   \centering
   \includegraphics[width=0.90\textwidth]{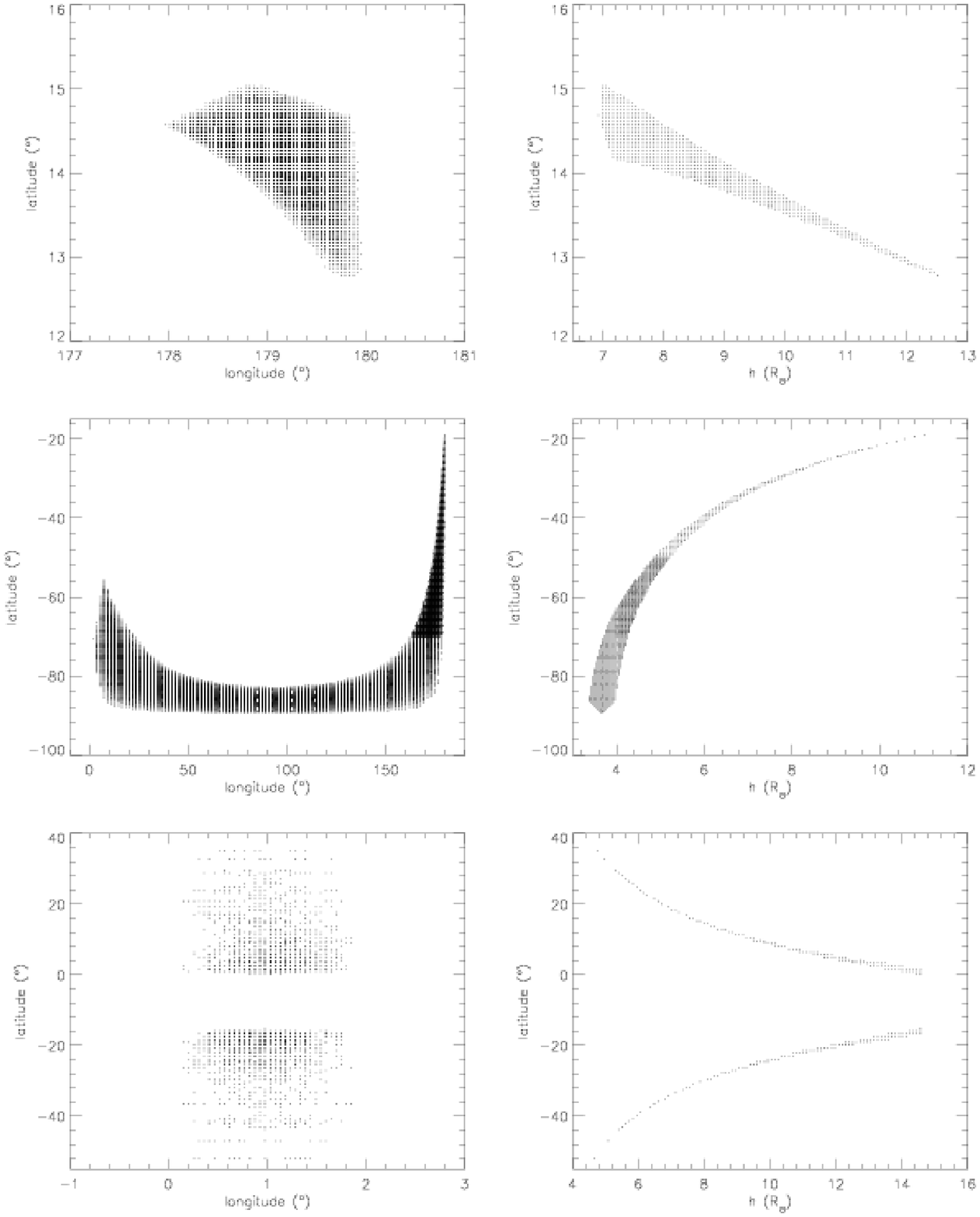}
   \caption{Results obtained for the flare observed in Algol with
     BeppoSAX for the simpler case of two contacts only (ingress and
     egress of the center of the emitting region). From top to
     bottom: solutions in the northern hemisphere (flare in the
     secondary eclipsed by the primary), solutions in the
     southern hemisphere, and solutions for the case of eclipse
     of the flare by the secondary star itself.}
   \label{fig:test1}
\end{figure*}
}

\onlfig{17}{
\begin{figure*}
   \centering
   \includegraphics[width=0.90\textwidth]{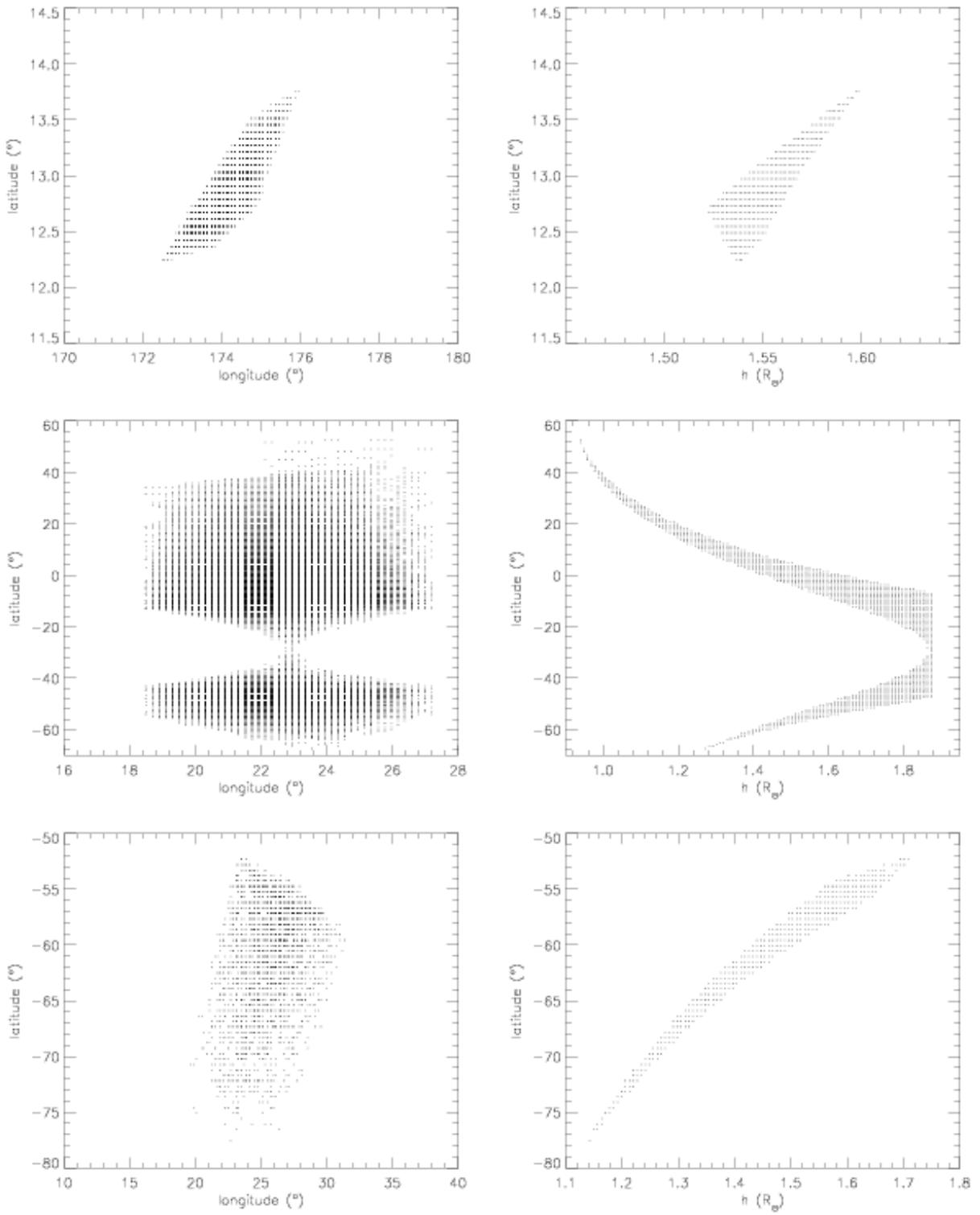}
   \caption{Results for the case of two contacts only in VW Cep. From
     top to bottom: flare in secondary eclipsed by primary,
     self-eclipsed by secondary only, and eclipse starting behind
     secondary and ending behind primary.}
   \label{fig:test2}
\end{figure*}
}

\onlfig{18}{
\begin{figure*}
   \centering
   \includegraphics[width=0.90\textwidth]{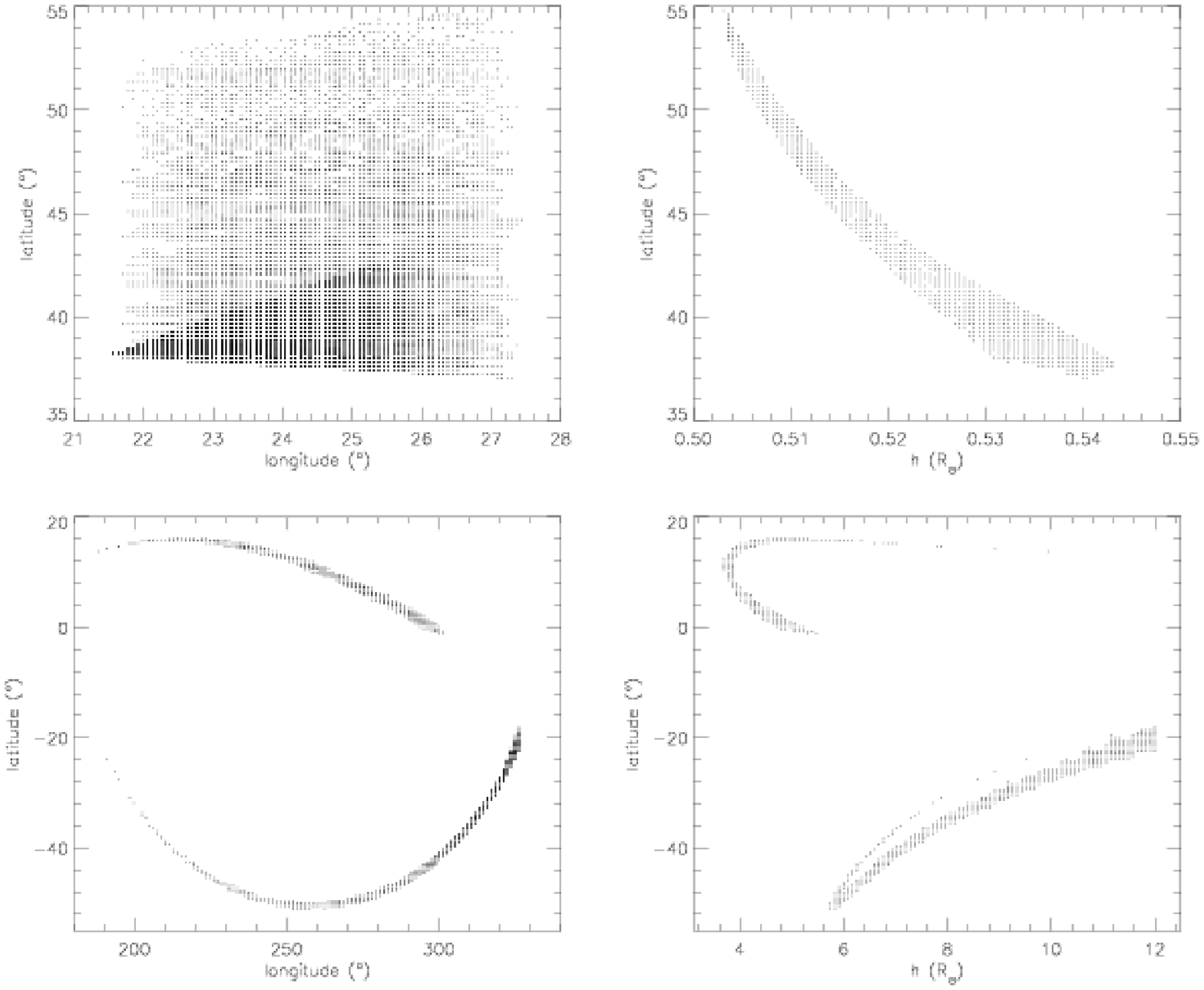}
   \caption{Results for the case of two contacts only. {\it Upper} panels:
     VW Cep, self-eclipsed by primary. {\it Lower} panels: eclipse in Algol
     observed by XMM (flare in secondary, eclipsed by primary).}
   \label{fig:test3}
\end{figure*}
}

While the original analyses of these events pointed towards polar
flares, \citep{cho98,schm99}, this work has 
shown that other configurations are possible from the geometric point of view.
A polar location is only possible for the Algol flare observed with
{\em BeppoSAX}.  
In this work we do not privilege any solution
among the possible results, and consider all configurations as equally
possible as 
long as they do not violate any physical law. Gravity forces or mass
transfer between close binaries should be considered in order to
constrain the number of solutions. 

The analysis poses a clear lower limit in electron density of $\log
n_e$\,(cm$^{-3}$)=10.4, well in agreement with the minimum values
determined from line ratios analysis in active stars. The largest
values compatible with the 
geometrical configurations, specially those with $\log
n_e$\,(cm$^{-3}$)\,$\ga 13$, are 
larger than those commonly observed in the X-ray line ratios of most
stars \citep[e.g.][]{sanz03,tes04,nes04}. The measurements of
the size of the emitting region shows that although the emitting
regions are small compared to the stellar size, we cannot reject
large loop lengths (measured through the variable $h$) by applying
only geometrical constrains based on the eclipse contacts.

\section{Conclusions}
We have systematically searched the solution space to calculate the geometrical
characteristics of the emitting region of four flares in active
stars. The method uses the times of the four contacts of the eclipse
and simulates all possible geometrical situations assuming that the
emitting region has a spherical shape and no substantial geometrical
changes during the eclipse. Two tests (including the use of a
  loop) were conducted to prove that the
approximation of a spherical shape is appropriate, even if the real
shape is not a sphere. 
The solutions found in the three flares
analyzed in this work and that of Paper I show that polar
flares are not possible in three of the four flares, and electron
densities must be larger than  $\log n_e$\,(cm$^{-3}$)=10.4,
consistent with measurements from line ratios of density sensitive
lines in active stars. The magnetic fields needed to confine these
loops range in both 
systems between
75~G and 3300~G.  The emitting regions have sizes that range from 0.002 to
0.5~$R_*$, and they can be at a distance (interpreted as loop
length) of up to 3.1~$R_*$ from the star's surface. 
Further refinement based on other physical limitations, such as gravity
or mass transfer could further constrain the set of solutions. 

\begin{acknowledgements}
This research is based on observations obtained with {\em XMM-Newton}, an
ESA science mission with instruments and contributions directly funded
by ESA Member States and NASA.
This research has made use
of NASA's Astrophysics Data System Abstract Service.
JS acknowledges the support by the ESA Research Fellowship Program. 
\end{acknowledgements}



\begin{thebibliography}{10}
\expandafter\ifx\csname natexlab\endcsname\relax\def\natexlab#1{#1}\fi

\bibitem[{{Al-Naimiy} {et~al.}(1985){Al-Naimiy}, {Mutter}, \& {Flaih}}]{aln85}
{Al-Naimiy}, H.~M.~K., {Mutter}, A.~A.~A., \& {Flaih}, H.~A. 1985, \apss, 108,
  227

\bibitem[{{Aluigi} {et~al.}(1994){Aluigi}, {Galli}, \& {Gaspani}}]{alu94}
{Aluigi}, M., {Galli}, G., \& {Gaspani}, A. 1994, Informational Bulletin on
  Variable Stars, 4117, 1

\bibitem[{{Choi} \& {Dotani}(1998)}]{cho98}
{Choi}, C.~S. \& {Dotani}, T. 1998, \apj, 492, 761

\bibitem[{{Hill}(1989)}]{hil89}
{Hill}, G. 1989, \aap, 218, 141

\bibitem[{{Ness} {et~al.}(2004){Ness}, {G{\" u}del}, {Schmitt}, {Audard}, \&
  {Telleschi}}]{nes04}
{Ness}, J.-U., {G{\" u}del}, M., {Schmitt}, J.~H.~M.~M., {Audard}, M., \&
  {Telleschi}, A. 2004, \aap, 427, 667

\bibitem[{{Sanz-Forcada} {et~al.}(2006){Sanz-Forcada}, {Favata}, \&
  {Micela}}]{sanz06}
{Sanz-Forcada}, J., {Favata}, F., \& {Micela}, G. 2006, \aap, 445, 673

\bibitem[{{Sanz-Forcada} {et~al.}(2003){Sanz-Forcada}, {Maggio}, \&
  {Micela}}]{sanz03}
{Sanz-Forcada}, J., {Maggio}, A., \& {Micela}, G. 2003, \aap, 408, 1087

\bibitem[{{Schmitt} \& {Favata}(1999)}]{schm99}
{Schmitt}, J. H. M.~M. \& {Favata}, F. 1999, \nat, 401, 44

\bibitem[{{Schmitt} {et~al.}(2003){Schmitt}, {Ness}, \& {Franco}}]{sch03}
{Schmitt}, J.~H.~M.~M., {Ness}, J.-U., \& {Franco}, G. 2003, \aap, 412, 849

\bibitem[{{Testa} {et~al.}(2004){Testa}, {Drake}, \& {Peres}}]{tes04}
{Testa}, P., {Drake}, J.~J., \& {Peres}, G. 2004, \apj, 617, 508

\end{thebibliography}
\end{document}